\documentclass[12pt]{article}
\usepackage{amsmath}
\usepackage{graphicx}
\usepackage{natbib}
\usepackage{url} 
\usepackage{tikz}
\usepackage{multirow}

\newcommand{\blind}{0}
\newcommand{\bx}{\textbf{x}}
\newcommand{\by}{\textbf{y}}
\newcommand{\be}{\textbf{e}}
\newcommand{\bX}{\textbf{X}}
\newcommand{\bY}{\textbf{Y}}
\newcommand{\bE}{\textbf{E}}
\newcommand{\bW}{\textbf{W}}
\newcommand{\ba}{\textbf{a}}
\newcommand{\bd}{\textbf{d}}
\newcommand{\bw}{\textbf{w}}
\newcommand{\low}{\mathcal{G}}
\newcommand{\high}{\mathcal{H}}
\newcommand{\bbeta}{\boldsymbol{\beta}}
\newcommand{\bSigma}{\boldsymbol{\Sigma}}
\newcommand{\rest}{\text{rest}}

\begin{document}

\def\spacingset#1{\renewcommand{\baselinestretch}%
{#1}\small\normalsize} \spacingset{1}


\if0\blind
{
  \title{\bf Bayesian Wavelet-packet Historical Functional Linear Models}
  \author{Mark J. Meyer\thanks{
    This work was supported by grants from the National Institutes of Health (ES007142, ES000002, ES016454, CA134294). The authors would like to thank Dr. David C. Christiani for use of the journeyman boilermaker data.}\hspace{.2cm}\\
    Department of Mathematics and Statistics, Georgetown University\\
    and \\
    Elizabeth J. Malloy \\
    Department of Mathematics and Statistics, American University\\
    and\\
    Brent A. Coull\\
    Department of Biostatistics, Harvard T. H. Chan School of Public Health}
  \maketitle
} \fi

\if1\blind
{
  \bigskip
  \bigskip
  \bigskip
  \begin{center}
    {\LARGE\bf Bayesian Wavelet-packet Historical Functional Linear Models}
\end{center}
  \medskip
} \fi

\bigskip
\begin{abstract}
Historical Functional Linear Models (HFLM) quantify associations between a functional predictor and functional outcome where the predictor is an exposure variable that occurs before, or at least concurrently with, the outcome. Current work on the HFLM is largely limited to frequentist estimation techniques that employ spline-based basis representations. In this work, we propose a novel use of the discrete wavelet-packet transformation, which has not previously been used in functional models, to estimate historical relationships in a fully Bayesian model. Since inference has not been an emphasis of the existing work on HFLMs, we also employ two established Bayesian inference procedures in this historical functional setting. We investigate the operating characteristics of our wavelet-packet HFLM, as well as the two inference procedures, in simulation and use the model to analyze data on the impact of lagged exposure to particulate matter finer than 2.5$\mu$g on heart rate variability in a cohort of journeyman boilermakers over the course of a day's shift.
\end{abstract}

\noindent%
\vfill

\newpage
\spacingset{1.5} 
\section{Introduction}
\label{s:intro}

Historical Functional Linear Models (HFLMs) are used to analyze the relationship between a functional ``exposure'' and a functional ``outcome'' where only exposures occurring in time before or concurrently with the outcome can affect the outcome. HFLMs are a special case of the Function-on-Function Regression (FFR) model which fits an unconstrained surface and is therefore inappropriate for modeling functional predictors that are lagged exposures. For example, suppose that for subject $i$, $x_i(v)$ represents levels of a pollutant sampled on a grid $v \in \mathcal{V}$ and $y_i(t)$ represents measurements of heart rate variability (HRV) sampled on a grid $t \in \mathcal{T}$. A general FFR model with no constraints  takes the form
	\begin{align}
		y_{i}(t) = \alpha(t) + \int_{v \in \mathcal{V}} x_{i}(v)\beta(v,t) dv + E_{i}(t),
		\label{eq:ffr}
	\end{align}
where the surface $\beta(v,t)$ is the primary quantity of interest for estimation and $E_{i}(t)$ is typically assumed to be distributed as a Gaussian Process. For example, see \cite{Ivanescu2015}, \cite{Meyer2015}, \cite{Morris2015}, \cite{Scheipl2015}, \cite{Scheipl2016}, \cite{Kim2018}, and references therein.

However, when applied to the HRV and pollutant example, a model like \eqref{eq:ffr} would allow for HRV measurements at time $t$ to be associated with  pollutants occurring both prior to time $t$, as well as after time $t$ despite the implausibility of such a relationship. The HFLM addresses this issue, constraining $\beta(v,t)$ to prevent such spurious associations by limiting the integration in \eqref{eq:ffr} to the set of coefficients such that $\{v \in \mathcal{V}, t \in \mathcal{T} : v \leq t\}$. The basic HFLM takes the form
	\begin{align}
		y_{i}(t) = \alpha(t) + \int_{\{v \leq t\}} x_{i}(v)\beta(v,t) dv + E_{i}(t).
		\label{eq:hflm}
	\end{align}
The observed data is usually discrete, so that $\beta(v,t)$ can be expressed as a matrix of coefficients. Thus the problem reduces to constraining the estimate of $\beta(v,t)$ to be non-zero for the upper triangle of the matrix.

Several authors explore ways of implementing the constraint in \eqref{eq:hflm}. \cite{MalfaitRamsay2003} propose the use tent-shaped basis functions with support over a two-dimensional region. They estimate the surface using a multivariate linear model approximation to a finite dimensional model. Thus after dimension reduction via the basis-space expansion, they use least squares to estimate $\beta(v,t)$. \cite{Harezlak2007} also use basis functions defined over a two-dimensional region but specify a large number of basis functions and penalize the fit. The authors consider both the LASSO and $L_2$-norm penalty on triangular basis functions, using restricted maximum likelihood (REML) for the latter. Both \cite{MalfaitRamsay2003} and \cite{Harezlak2007} allow for a pre-defined lag beyond which the effect of exposure is zero, further constraining the surface to a trapezoidal region defined by $\{v, \in \mathcal{V}, t \in \mathcal{T} : t - \Delta \leq v \leq t\}$ for some pre-defined lag $\Delta$. \cite{Kim2011} take the constraint further by proposing a Recent History Functional Linear Model where the surface is constrained to a trapezoidal region defined by $\{v, \in \mathcal{V}, t \in \mathcal{T} : t - \Delta_1 \leq v \leq t - \Delta_2\}$ for $0 < \Delta_1 < \Delta_2 < T$. The authors estimate the constrained surface with a varying coefficient model representation using B-spline basis functions, although they suggest Fourier, truncated power, and Eigen basis functions can also be used.

More recently, \cite{Pomann2016} examine two HFLMs that allow for multiple functional predictors with estimation constrained to a fixed window similar to that used in \cite{Harezlak2007}. The authors implement two approaches to estimation using semi-local smoothing, which performs point-wise estimation, and global smoothing, which smooths over $\mathcal{T}$ globally. The methods select smoothing parameters via cross-validation and REML, respectively, and use B-spline basis expansions to model the functional form. In each of these methods, the covariance of the error term, $E_i(t)$, is assumed independent or to have ``working'' independence.

To the best of our knowledge, the existing body of work on HFLMs is largely limited to spline-based methods which can over smooth signals and peaks in spiky and irregular data. Further, the current literature does not consider inferential procedures, focusing on estimation and model fit criterion instead. As such, the performance of the proposed methods with respect to uncertainty quantification is not clear. Wavelet-based functional regression models, such as the work of \cite{MorrisCarroll2006} and \cite{Malloy2010}, consider the function-on-scalar and scalar-on-function regression settings, respectively, in the Bayesian context. \cite{Meyer2015} extend \cite{MorrisCarroll2006} to the FFR case using wavelets for the basis function of the outcome and wavelet-principal components for the expansion of the predictor. One advantage of the wavelet-based framework is that it does not require the assumption of independence in the data-space. However, their approach estimates an unconstrained surface and therefore is inappropriate for modeling lagged exposures. The wavelet domain \cite{Meyer2015} use, which results from a discrete wavelet transformation, lacks a convenient relationship between the wavelet coefficients and the time domain, thus making it inefficient for use in an HFLM. Furthermore, the wavelet-principal components basis function does not preserve the temporal relationship between exposure and outcome in the wavelet domain. Thus to implement an HFLM using wavelets, a different basis function is warranted. Wavelet-packets, which result from a discrete wavelet-packet transformation, are a variant of the wavelet basis function that have a convenient relationship to the time domain that we can exploit to implement a wavelet-packet based HFLM.

In this work, we propose a novel use of wavelet-packets that allows us to build a Bayesian wavelet-space HFLM. We formulate our model within the framework of \cite{MorrisCarroll2006} and \cite{Meyer2015}, thus our method does not require the assumption of independence in the data space. A benefit of the Bayesian context is that we can implement several Bayesian techniques for conducting multiplicity adjusted inference including joint credible intervals to quantify uncertainty and the Bayesian false discovery rate to identify exposure lags and exposure times that are associated with the outcome. We assess the operating characteristics of the methodology and inference procedures in simulation and present an application to data from a study of journeyman boilermakers exposed to varying levels of particulate matter during the course of the day. The data consists of five-minute assessments of HRV, as defined by standard deviation of the normal-to-normal intervals (SDNN) at each time-point, and particulate matter finer than 2.5 $\mu$m (PM$_{2.5}$) which results from exposure to residual oil fly ash and cigarette smoke \citep{Magari2001,Cavallari2008}. \cite{Harezlak2007} present an analysis of part of this data and found both negative and positive time-specific associations in the morning that corresponded to the workers' break times. Our analysis focuses on the morning hours of the work day where \cite{Harezlak2007} saw the largest effects. We also make available MATLAB code for the implementation of our method at \url{https://github.com/markjmeyer/WPHFLM}.

The remainder of the paper is organized as follows: Section~\ref{s:dwpt} provides a brief introduction to the discrete wavelet-packet transformation. Section~\ref{s:BHFLM} details the formulation of the the Bayesian 
wavelet-packet HFLM along with a discussion of inferential procedures in this modeling framework. Sections~\ref{s:sim} and~\ref{s:app} present the results of our simulation study and the application of our model to the Journeyman data, respectively. Finally, in Section~\ref{s:disc}, we provide a discussion of the methodology.

\section{Discrete Wavelet-packet Transformation}
\label{s:dwpt}

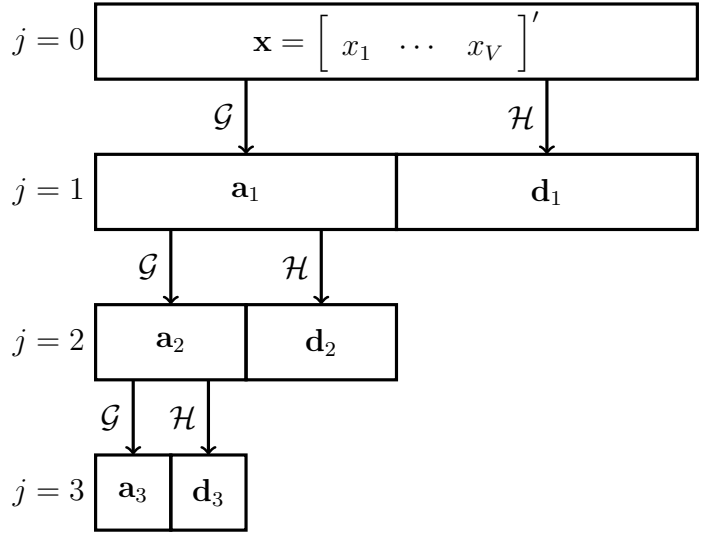
\begin{figure}
	\begin{center}
	    	\begin{tikzpicture}
	    		\draw[very thick] (0,0) rectangle (8,1) node[midway] {$\bx = \left[ \begin{array}{ccc} x_1 & \cdots & x_V \end{array} \right]'$};
			\node[left] at (0,0.5) {$j = 0$};

			
			\draw[very thick, ->] (2,0) -- (2,-1) node[midway, left]  {$\low$};
			\draw[very thick, ->] (6,0) -- (6,-1) node[midway, left]  {$\high$};

	    		\draw[very thick] (0,-2) rectangle (4,-1) node[midway] {$\ba_1$};
	    		\draw[very thick] (4,-2) rectangle (8,-1) node[midway] {$\bd_1$};
			\node[left] at (0,-1.5) {$j = 1$};

			\draw[very thick, ->] (1,-2) -- (1,-3) node[midway, left]  {$\low$};
			\draw[very thick, ->] (3,-2) -- (3,-3)  node[midway, left]  {$\high$};

	    		\draw[very thick] (0,-4) rectangle (2,-3) node[midway] {$\ba_2$};
	    		\draw[very thick] (2,-4) rectangle (4,-3) node[midway] {$\bd_2$};
			\node[left] at (0,-3.5) {$j = 2$};

			\draw[very thick, ->] (0.5,-4) -- (0.5,-5) node[midway, left]  {$\low$};
			\draw[very thick, ->] (1.5,-4) -- (1.5,-5)  node[midway, left]  {$\high$};

	    		\draw[very thick] (0,-6) rectangle (1,-5) node[midway] {$\ba_3$};
	    		\draw[very thick] (1,-6) rectangle (2,-5) node[midway] {$\bd_3$};
			\node[left] at (0,-5.5) {$j = 3$};
	    	\end{tikzpicture}
		\caption{\label{f:dwt}Pyramid algorithm for partial DWT \citep{PercivalWalden2000}.}
	\end{center}
\end{figure}

We begin with a brief description of the partial discrete wavelet transformation (DWT). Consider a 1-dimensional function, $x(v)$, which we discretely observe as $\bx = \left[ \begin{array}{ccc} x_1 & \cdots & x_V \end{array} \right]'$, where $V = 2^k$ for some positive integer value $k$. For a given mother wavelet, a partial DWT to $J = 3$ levels results in a set of wavelet coefficients that can be further separated into approximation coefficients, $\ba$, and detail coefficients, $\bd$. The pyramid algorithm for performing the partial DWT is graphically depicted in Figure~\ref{f:dwt}. Here, $\low$ and $\high$ denote the low and high pass filters for the corresponding mother wavelet. The resulting decomposition consists of the detail coefficients from each level plus the final approximation coefficients. Post-transformation, the wavelet-space representation of $\bx$ is then $\bw = \left[ \begin{array}{cccc} \ba_3 & \bd_3 & \bd_2 & \bd_1 \end{array} \right]'$. The DWT can be expressed as a matrix multiplication by an orthogonal matrix, $W$, thus it can be shown that $\bw = \bx W$.

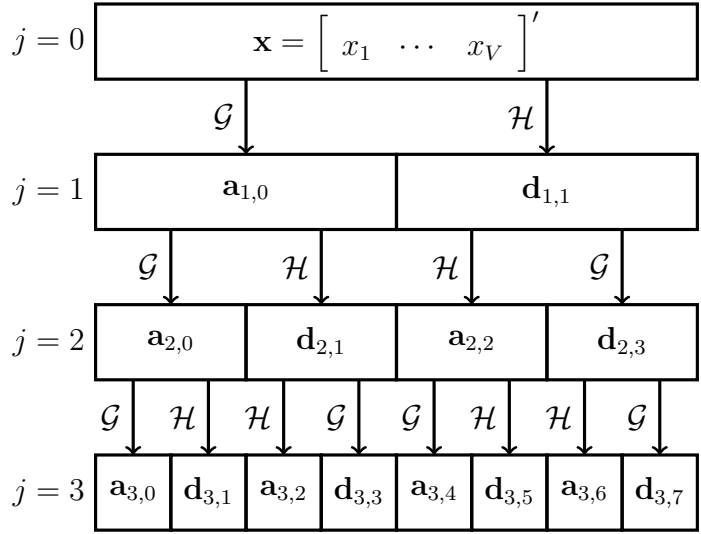
\begin{figure}
	\begin{center}
	    	\begin{tikzpicture}
	    		\draw[very thick] (0,0) rectangle (8,1) node[midway] {$\bx = \left[ \begin{array}{ccc} x_1 & \cdots & x_V \end{array} \right]'$};
			\node[left] at (0,0.5) {$j = 0$};
			
			\draw[very thick, ->] (2,0) -- (2,-1) node[midway, left]  {$\low$};
			\draw[very thick, ->] (6,0) -- (6,-1) node[midway, left]  {$\high$};

	    		\draw[very thick] (0,-2) rectangle (4,-1) node[midway] {$\ba_{1,0}$};
	    		\draw[very thick] (4,-2) rectangle (8,-1) node[midway] {$\bd_{1,1}$};
			\node[left] at (0,-1.5) {$j = 1$};

			\draw[very thick, ->] (1,-2) -- (1,-3) node[midway, left]  {$\low$};
			\draw[very thick, ->] (3,-2) -- (3,-3)  node[midway, left]  {$\high$};
			\draw[very thick, ->] (5,-2) -- (5,-3) node[midway, left]  {$\high$};
			\draw[very thick, ->] (7,-2) -- (7,-3)  node[midway, left]  {$\low$};

	    		\draw[very thick] (0,-4) rectangle (2,-3) node[midway] {$\ba_{2,0}$};
	    		\draw[very thick] (2,-4) rectangle (4,-3) node[midway] {$\bd_{2,1}$};
	    		\draw[very thick] (4,-4) rectangle (6,-3) node[midway] {$\ba_{2,2}$};
	    		\draw[very thick] (6,-4) rectangle (8,-3) node[midway] {$\bd_{2,3}$};
			\node[left] at (0,-3.5) {$j = 2$};

			\draw[very thick, ->] (0.5,-4) -- (0.5,-5) node[midway, left]  {$\low$};
			\draw[very thick, ->] (1.5,-4) -- (1.5,-5)  node[midway, left]  {$\high$};
			\draw[very thick, ->] (2.5,-4) -- (2.5,-5) node[midway, left]  {$\high$};
			\draw[very thick, ->] (3.5,-4) -- (3.5,-5)  node[midway, left]  {$\low$};
			\draw[very thick, ->] (4.5,-4) -- (4.5,-5) node[midway, left]  {$\low$};
			\draw[very thick, ->] (5.5,-4) -- (5.5,-5)  node[midway, left]  {$\high$};
			\draw[very thick, ->] (6.5,-4) -- (6.5,-5) node[midway, left]  {$\high$};
			\draw[very thick, ->] (7.5,-4) -- (7.5,-5)  node[midway, left]  {$\low$};

	    		\draw[very thick] (0,-6) rectangle (1,-5) node[midway] {$\ba_{3,0}$};
	    		\draw[very thick] (1,-6) rectangle (2,-5) node[midway] {$\bd_{3,1}$};
	    		\draw[very thick] (2,-6) rectangle (3,-5) node[midway] {$\ba_{3,2}$};
	    		\draw[very thick] (3,-6) rectangle (4,-5) node[midway] {$\bd_{3,3}$};
	    		\draw[very thick] (4,-6) rectangle (5,-5) node[midway] {$\ba_{3,4}$};
	    		\draw[very thick] (5,-6) rectangle (6,-5) node[midway] {$\bd_{3,5}$};
	    		\draw[very thick] (6,-6) rectangle (7,-5) node[midway] {$\ba_{3,6}$};
	    		\draw[very thick] (7,-6) rectangle (8,-5) node[midway] {$\bd_{3,7}$};
			\node[left] at (0,-5.5) {$j = 3$};
	    	\end{tikzpicture}
		\caption{\label{f:dwpt}Algorithm for partial DWPT \citep{PercivalWalden2000}.}
	\end{center}
\end{figure}

From Figure~\ref{f:dwt}, we see the DWT applies the filters only to the successive sets of approximation coefficients. For a given mother wavelet, the partial discrete wavelet-packet transformation (DWPT) begins in the same way, passing first the low and then high-pass filters over the original signal. However, after the first level, the DWPT applies $\low$ and $\high$ to all coefficients at each level, reversing the order when applied to detail coefficients. Figure~\ref{f:dwpt} illustrates the partial DWPT decomposition for $J=3$ levels. Once again, $\low$ and $\high$ denote the low and high pass filters of the corresponding mother wavelet. The resulting representation of $\bx$ is then the final set of coefficients from the last level, $\bw_P = \left[ \begin{array}{cccccccc} \ba_{3,0} & \bd_{3,1} & \ba_{3,2} & \bd_{3,3} & \ba_{3,4} & \bd_{3,5} & \ba_{3,6} & \bd_{3,7} \end{array} \right]'$. As in the DWT, the DWPT can also be expressed as a matrix multiplication of an orthogonal matrix, $W_P$. Thus $\bw_P$ can be shown to be $\bw_P = \bx W_P$. For both the DWT and DWPT, the resulting set of wavelet-packet coefficients will have the same length as the original signal.

Similar to the DWT, the wavelet-packet coefficients comprising $\bw_P$ are indexed by a scale and location. The scale indexes the bin the coefficient is in at the final level of the decomposition and thus corresponds to the second subscript from the $j = 3$ level in Figure~\ref{f:dwpt} while the location denotes the position of the coefficient within the set of coefficients at a given scale. Since the number of elements in $\bw_P$ is the same as in $\bx$, within a scale, the time-ordering of the observed signal is preserved in the ordering of the coefficients. Suppose we have a second function, $y(t)$, that we discretely observe on a grid such that $\by = \left[ \begin{array}{ccc} y_1 & \cdots & y_T \end{array} \right]'$. If we perform the DWPT on $\by$, we will also obtain a set of wavelet-packet coefficients that preserve the time-ordering of the original signal within each scale. Provided the elements of $\bx$ are sampled in time concurrently or before the elements of $\by$, we can use the location index from their respective DWPT decompositions to constrain the surface of estimation within each scale in the wavelet-packet space. When we apply the inverse DWPT (IDWPT), the constraint in then maintained in the data-space. For more details on DWPTs, see \citet[chap. 6]{PercivalWalden2000}, \cite{Misiti2007}, and \citet[chap. 2]{Nason2008}.

\section{Bayesian Historical Functional Linear Model}

\label{s:BHFLM}

We begin with the model in \eqref{eq:hflm} where we constrain the estimation to the region defined by $\{v, \in \mathcal{V}, t \in \mathcal{T} : v \leq t\}$. We assume the within-function errors come from a Gaussian process. Thus, $E_{i}(t) \sim \mathcal{GP}\left(0, \Sigma_E\right)$, where $\Sigma_E$ is an unstructured covariance matrix. Because the data, $y_{i}(t)$ and $x_i(v)$, arrive sampled on a grid of equally spaced time points $t = [t_1,\cdots,t_T]'$ and  $v = [v_1,\cdots,v_V]'$, we use the discrete version of the model: $\by_{i} = \bx_{i}\bbeta + \be_{i}$ for the vectors $\by_{i}$, $\bx_i$, and $\be_i$ and matrix of coefficients $\bbeta$. We recommend centering and scaling both the outcome and predictor functions first. Thus, without loss of generality, we drop the intercept function from the model. Stacking the response vectors and predictor vectors into matrices gives
	\begin{equation}
		\bY= \bX\bbeta + \bE,
		\label{eq:dhflm}
	\end{equation}
where for $N$ total curves, $\bE$ and $\bY$ are $N\times T$ while $\bX$ is $N\times V$.
The constrained region of integration in Model \eqref{eq:hflm} restricts the form of the functional regression coefficients so that $\beta(v_k,t_{k'}) = 0$ if $v_k > t_{k'}$. If $T = V$ and $t_1 = v_1, t_2 = v_2, \ldots, t_T = v_V$, then the discrete version of $\bbeta$ is an upper triangular matrix of the form
	\begin{equation}
		\bbeta = \left(\begin{array}{cccc}
			\beta(v_1,t_1) & \beta(v_1,t_2) & \cdots & \beta(v_1,t_T) \\
			0 & \beta(v_2,t_2) & \cdots & \beta(v_2,t_T) \\
			\vdots & \vdots & \ddots & \vdots \\
			0 & 0 & \cdots & \beta(v_V,t_T)
			\end{array} \right)
		\label{eq:hbeta}
	\end{equation}
with zeros below the main diagonal. We propose the use of wavelet-packets to enforce this constraint in the wavelet-packet space which, given that the time ordering is preserved, will ensure the constraint is maintained in the data-space as well.

\subsection{Historical Constraint via Wavelet-Packets}

	Working with \eqref{eq:dhflm}, we apply the DWPT separately to each row of $\bY$ and to each row of $\bX$. Performing this transformation is equivalent to the post-multiplication of the approximately orthonormal projection matrices resulting from the DWPT \citep{PercivalWalden2000}. The resulting decompositions have the form $\textbf{Y} = \textbf{Y}^{W_P} \bW_{P,Y}$ and $\textbf{X} = \textbf{X}^{W_P} \bW_{P,X}$ where $\bW_{P,Y}$ and $\bW_{P,X}$ are orthogonal matrices containing the wavelet packet basis functions. Then for the two dimensional decomposition on $\boldsymbol{\beta} = \bW_{P,X}' \boldsymbol{\beta}^{W_P} \bW_{P,Y}$ Model~(\ref{eq:dhflm}) in the wavelet-packet space is
		$\bY^{W_P} \bW_{P,Y} = \bX^{W_P}\bW_{P,X} \bW_{P,X}'\boldsymbol{\beta}^{W_P}\bW_{P,Y} +  \bE^{W_P}\bW_{P,Y}$
for $\bE = \bE^{W_P}\bW_{P,Y}$. Post-multiplying by $W_{P,Y}$ and recognizing the orthogonality of the wavelet-packet basis matrices, this model reduces to $\bY^{W_P} = \bX^{W_P}\boldsymbol{\beta}^{W_P} +  \bE^{W_P}$ and subject-specific model $\by_i^{W_P} = \bx_i^{W_P}\boldsymbol{\beta}^{W_P} +  \be_i^{W_P}$. This model fits into the generalized basis expansion framework for function-on-function regression considered in \cite{Meyer2015}. The key difference, however, lies in our novel use of wavelet-packets as a basis function to induce the historical constraint whereas \cite{Meyer2015} is concerned only with basis functions to estimate a full, unconstrained surface or set of surfaces.

We enforce the constraint in the wavelet-packet space via our prior specification on the elements of $\bbeta$.
Let the DWP transformations be indexed by scales $j = 1, \ldots, J^y$ and $s = 1, \ldots, S^x$ and locations $k = 1, \ldots, K_j^y$ and $\ell = 1, \ldots, L_s^x$ in the $\bY$ and $\bX$ wavelet-packet spaces respectively. Consistent with previous work on wavelet-based models in function regression, we place spike-and-slab priors on model coefficients. To restrict the surface in wavelet-packet space, we set coefficients where $\ell > k$ to zero. Thus our prior on the elements of $\boldsymbol{\beta}^{W_P} = \left[\beta^{W_P}_{s\ell,jk}\right]$ is
		$\beta_{s\ell,jk}^{W_P} \sim\ 1(\ell \leq k)\gamma_{s\ell,jk}N(0, \tau_{s\ell,j\cdot}) + [1-\gamma_{s\ell,jk}]d_0$, where $\gamma_{s\ell,jk} \sim Bern(\pi_{s\ell,j\cdot})$
and $d_0$ is a point-mass distribution at zero. The regularization parameters, $\tau_{s\ell,j\cdot}$ and $\pi_{s\ell,j\cdot}$, smooth over locations $k$ which we denote using the ``dot'' notation in the subscript. We assume inverse-gamma and beta hyper priors, respectively, for the regularization parameters with hyper-parameters fixed and based on the data.

\cite{MorrisCarroll2006} show that after a wavelet transformation, assuming independence in the wavelet-space does not imply independence in the data-space and therefore wavelets accommodate a wide range of covariances in the data-space. As wavelet-packets share the same whitening properties of wavelets, we assume independence in the wavelet-packet space \citep{PercivalWalden2000}. Thus we assume $\be_i^{W_p} \sim N(0, \bSigma^{W_p})$ where $\bSigma^{W_p} = \text{diag}\left\{ \sigma^2_{jk} \right\}$, which varies by the scale and location of the $\bY$ wavelet-packet coefficients. We place an inverse gamma prior on $\sigma^2_{jk}$. The independence assumption allows us to sample the coefficients corresponding to different $j$ and $k$ combinations separately.

Using the prior specifications, we now describe our sampling algorithm. For the $jk$th wavelet-packet space coefficient from the $\bY^{W_P}$ decomposition and the $s\ell$th column of $\bX^{W_P}$, the conditional posterior distribution is a mixture of a point-mass at zero and a normal distribution of the form
\begin{align}
	\beta^{W_P}_{s\ell,jk} | \rest \sim 1(\ell \leq k)\gamma_{s\ell,jk} N(\mu_{s\ell,jk}, \epsilon_{s\ell,jk}) + (1 - \gamma_{s\ell,jk})d_0, \label{eq:beta}
\end{align}
where $\mu_{s\ell,jk} = \hat{\beta}_{s\ell,jk}^{W_P}(1 + \Lambda_{s\ell,jk}/\tau_{s\ell,j})^{-1}$ and $\epsilon_{s\ell,jk} = \Lambda_{s\ell,jk}(1 + \Lambda_{s\ell,jk}/\tau_{s\ell,j})^{-1}$ for the OLS and variance estimates $\hat{\beta}_{s\ell,jk}^{W_P}$ and $\Lambda_{s\ell, jk}$ at the current step. The conditional for $\gamma_{s\ell,jk}$ is
\begin{align}
	\gamma_{s\ell,jk}|\rest \sim Bern(\alpha_{s\ell,jk}),\label{eq:gamma}
\end{align}
were $\alpha_{s\ell,jk} = O_{s\ell,jk}/\left( O_{s\ell,jk} + 1 \right)$ for $O_{s\ell,jk} = \pi_{s\ell,j}/(1-\pi_{s\ell,j})\text{BF}_{s\ell,jk}$, $\text{BF}_{s\ell,jk} = \left( 1 + \frac{\tau_{s\ell,j}}{\Lambda_{s\ell, jk}} \right)^{-1/2}$ $\exp{\left\{ \frac{1}{2} \zeta^2_{s\ell,jk}\left(1 + \frac{\Lambda_{s\ell, jk}}{\tau_{s\ell,j}}\right) \right\}}$, and $\zeta_{s\ell,jk}$ equal to the ratio of the current values of $\beta^{W_P}_{s\ell,jk}$ to the current estimate of the standard deviation of $\beta^{W_P}_{s\ell,jk}$. Through the indicator function, $1(\ell \leq k)$, we enforce the historical constraint by forcing coefficients for which $\ell > k$ to come from the point-mass density $d_0$ and thus be set to zero.

The conditionals for the diagonal elements of the wavelet-packet space variance components, $\sigma^2_{jk}$, have the form
\begin{align}
	P\left(\sigma^2_{jk} | \rest\right) &\propto \pi\left(\sigma^2_{jk} \right)\left( \sigma^2_{jk} \right)^{-n/2} \exp\left[ -\frac{1}{2\sigma^2_{jk}} \left(\by^{W_p}_{jk} - X\beta^{W_p}_{\cdot\cdot,jk}\right)'\left(\by^{W_p}_{jk} - X\beta^{W_p}_{\cdot\cdot,jk}\right) \right], \label{eq:sig}
\end{align}
where $\pi\left(\sigma^2_{jk} \right)$ is the prior density on $\sigma^2_{jk}$ which we take to be inverse gamma with parameters $a_{\sigma^2}$ and $b_{\sigma^2}$ both set to the empirical Bayes estimates. To ensure the variance components are not too close to zero, we employ a Metropolis-Hastings step to sample them. The proposal densities are independent Gaussians, truncated at zero and centered at the previous value in each chain. The full conditionals for the regulation parameters are
\begin{align}
	\tau_{s\ell,j\cdot} | \text{rest} &\sim IG\left[a_{\tau} + \frac{1}{2}\gamma_{s\ell,jk}, b_{\tau} + \frac{1}{2}\gamma_{s\ell,jk}\left(\beta_{s\ell,jk}^{W_P}\right)^2 \right] \text{ and }\label{eq:tau}\\
	\pi_{s\ell,j\cdot} | \text{rest} &\sim Beta\left( a_{\pi} + \gamma_{s\ell,jk}, b_{\pi} + \gamma_{s\ell,jk} \right),\label{eq:pi}
\end{align}
with $a_{\tau}$, $b_{\tau}$, $a_{\pi}$, and $b_{\pi}$ set to the empirical Bayes estimates. The prior specifications are consistent with the previous work on wavelet-based functional models including \cite{Meyer2015}, \cite{Malloy2010}, and \cite{MorrisCarroll2006}. For more details on the empirical Bayes estimates, see \cite{MorrisCarroll2006}. Our sampler then iterates between draws from \eqref{eq:beta} to \eqref{eq:pi} until convergence. Upon completion of the algorithm, we apply the inverse DWPT to the posterior samples of $\bbeta^{W_P}$ to obtain estimates in the data space of $\bbeta$, the upper triangular matrix of historically constrained coefficients.

\subsection{Thresholding and Wavelet Details}

Our sampler becomes computationally intensive as $V$ increases with computation time increasing almost linearly. In previous work on wavelet-based FFR, \cite{Meyer2015} address this issue by reducing the dimension of the wavelet transformed $\bX$. In their formulation, the authors use wavelet-Principal Components (wPC) retaining columns containing a large amount of the variability in $\bX$. The wPC decomposition involves first performing a DWT on $\bX$ and then performing a singular value decomposition (SVD). While such an approach reduces computation time and achieves additional denoising, it does not work in the historical framework, as performing an SVD on $\bX^{W_P}$ would break the time ordering maintained by the transformation. To remedy the computational concerns and simultaneously achieve additional denoising, we propose a simplified thresholding procedure where we threshold all coefficients in the larger scale levels to zero. Thus, we consider keeping the scales that comprise either 25\% or 50\% of the wavelet-packet coefficients. For example, using the $J = 3$ level partial DWPT, the first approach would result in retaining the coefficients $\ba_{3,0}$ and $\bd_{3,1}$ in Figure~\ref{f:dwpt}. The second approach, keeping 50\%, would correspond to retaining the coefficients $\ba_{3,0}$, $\bd_{3,1}$, $\ba_{3,2}$, and $\bd_{3,3}$, again from Figure~\ref{f:dwpt}. Our simulation setting compares the performance of both approaches in terms of computational efficiency and estimation. It is important to note that we do not threshold the outcome and retain all coefficients of $\bY^{W_P}$.

The choice of mother wavelet will depend largely on the data context. In our application and simulation, we use Daubechies wavelets with 3 vanishing moments for the separate DWPTs on both $\bY$ and $\bX$. Wavelet-packets also require a choice of boundary padding, we select zero padding---padding the boundary with zeros. Finally, we must choice the number of levels to take our partial DWPT. Because of our reliance on the maintained time ordering within scales but between decompositions, we use the same number of levels of decomposition for both $\bY$ and $\bX$. In simulation, we use $J = 3$ levels for sampling densities of $T = V = 64$ and $128$. For denser sampling rates, we recommend increasing $J$ and potentially adjusting the thresholding.

\subsection{Posterior Functional Inference}

Previous implementations of the HFLM do not discuss inferential procedures instead, in some cases, developing measures of model fit \citep{MalfaitRamsay2003,Harezlak2007}. In the wavelet-based functional literature, \cite{Meyer2015} propose the use of a Bayesian False Discovery Rate (BFDR) procedure also used by \cite{MorrisEtAl2008} and \cite{Malloy2010}, as well as joint credible bands as discussed in \cite{Ruppert2003}. We consider the use of both methods in the context of Bayesian wavelet-packet HFLM.

	The BFDR procedure utilizes the MCMC samples to estimate the posterior probability of a given coefficient being greater than a $\delta$-fold intensity change. Once these values are determined they are ranked and a cut-off selected to control the overall FDR at a pre-specified global $\alpha$-bound. Suppose we have $M$ MCMC samples and $\beta^{(m)}(v,t)$ is $m^{\text{th}}$ draw from the posterior estimated surface. Then for $\left\{ v\in \mathcal{V} \text{ and } t\in \mathcal{T} \text{ s.t. } v \leq t \right\}$, we find the probability
		$P_{B}(v,t) = Pr\left\{ \left| \beta(v,t) \right| > \delta | y \right\} \approx \frac{1}{M} \sum_{m=1}^{M} 1\left\{  \left| \beta^{(m)}(v,t) \right| > \delta \right\}.$
	We then flag the set of coefficients on the historical surface satisfying $\psi = \left\{ (v,t) : v \leq t \text{ and } P_{B}(v,t) \geq \phi_{\alpha} \right\}$. Here, $\phi_{\alpha}$ is determined by first ranking the values of $P_{B}$ in descending order across all coefficients to obtain the set $\left\{ P_{(r)}, r = 1,\ldots,R\right\}$ where $R$ is the total number of coefficients satisfying the historical constraint. We then define the cut-off value as $\lambda = \max\big[ r^* : \frac{1}{r^*} \sum_{r=1}^{r^*}\left\{ 1 - P_{(r)} \right\} \leq \alpha \big]$. Thus, we select coefficients with $P_B$ great than or equal to $\phi_{\alpha} = P_{(\lambda)}$ as significant.
	
	For interval estimation, we consider both point-wise credible intervals as well as joint credible intervals. Point-wise credible intervals are constructed by taking the $\alpha/2$ and $1 - \alpha/2$ quantiles of the posterior samples taken at each coefficient for some choice of $\alpha$. As in \cite{Meyer2015}, we construct joint credible intervals using $I_{\alpha}(v,t) = \hat{\beta}(v,t) \pm q_{(1-\alpha)}\left[ \widehat{\text{St.Dev}}\left\{ \hat{\beta}(v,t) \right\} \right]$, where $\hat{\beta}(v,t)$ and $\widehat{\text{St.Dev}}\left\{ \hat{\beta}(v,t) \right\}$ are the mean and standard deviation of the posterior samples, respectively, and $q_{(1-\alpha)}$ is the $(1-\alpha)$ quantile taken over all posterior samples of the quantity
	\begin{align*}
		q^{(m)} = \max_{(v,t)}\left| \frac{\beta^{(m)}(v,t) - \hat{\beta}(v,t)}{\widehat{\text{St.Dev}}\left\{ \hat{\beta}(v,t) \right\}} \right|, \text{ s.t. } v \leq t.
	\end{align*}
	Such an interval satisfies $Pr\left\{ L(v,t) \leq \beta(v,t) \leq U(v,t)\ \forall\ v \in \mathcal{V}, t \in \mathcal{T} \text{ s.t. } v \leq t \right\} \geq 1 - \alpha$, where $L(v,t)$ and $U(v,t)$ are the corresponding upper and lower interval bounds. This procedure yields a joint  $100(1-\alpha)$ interval for the historical association surface. 

\section{Simulation Study}
\label{s:sim}

We consider four different historical surface scenarios representing plausible relationships between $x(v)$ and $y(t)$: a lagged effect of $x(v)$ on $y(t)$, a cumulative effect, a time-specific effect, and a delayed time-specific effect. Figure~\ref{f:true} displays the surfaces. We present their mathematical expressions in the Supplementary Material. For each scenario, we assume the data is sampled on a grid between $t_1 = v_1 = 1$ and $t_T = v_V = 64$ and vary the sampling rate such that $T = V = 64$ or $T = V = 128$. We also vary the sample size, considering $N = 50$ and $N = 200$. For the sparser, $T = 64$ setting, we vary the percent of the columns of $\bX^{W_P}$ we retain, first retaining 25\% of coefficients and then retaining 50\%. In the $T = 128$ setting, we only retain 25\% of coefficients since retaining more considerably increases the computational burden. Thus in total, we present the results from 24 different simulation settings.

	\begin{figure}
		\centering
		\includegraphics[width = 3in]{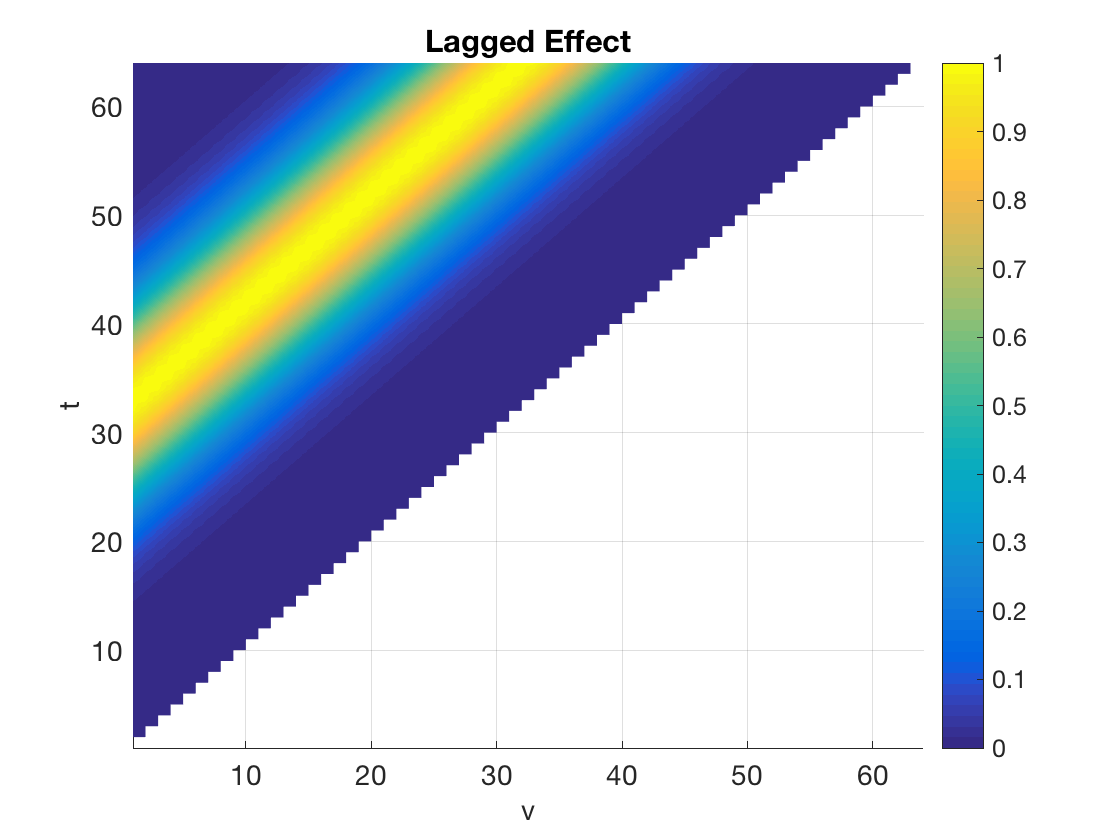}
		\includegraphics[width = 3in]{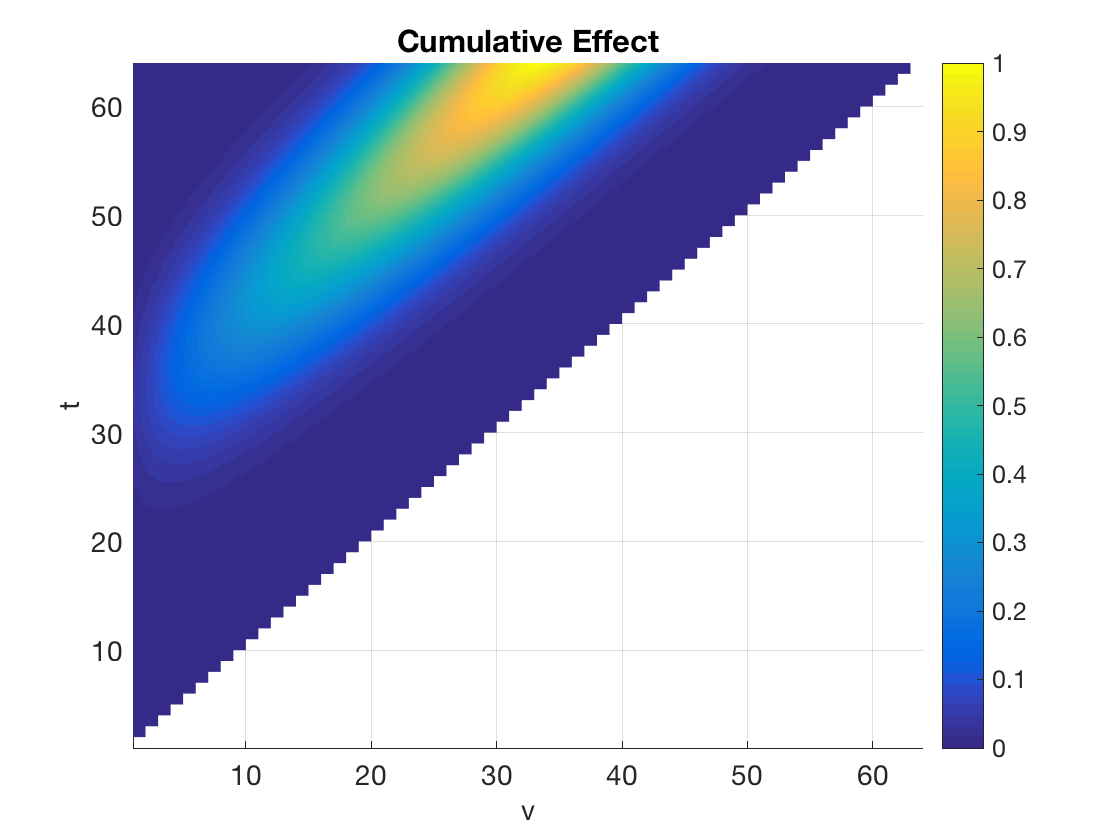}
		\includegraphics[width = 3in]{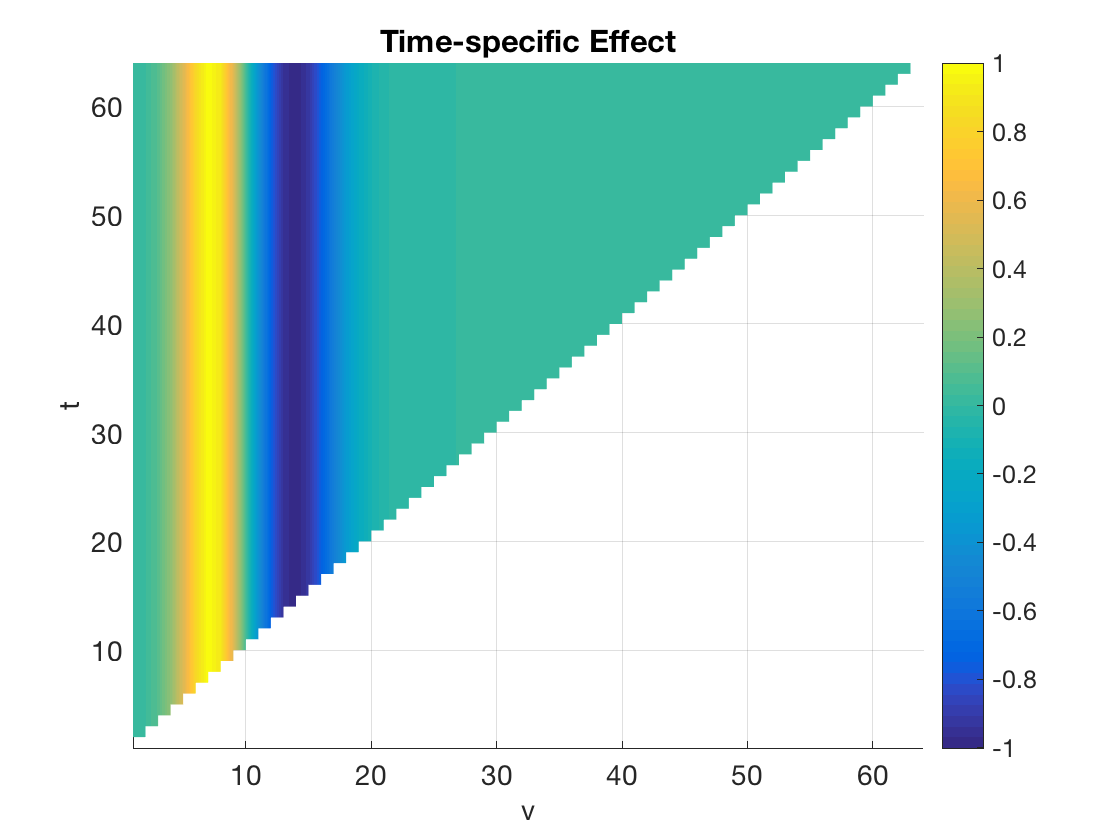}
		\includegraphics[width = 3in]{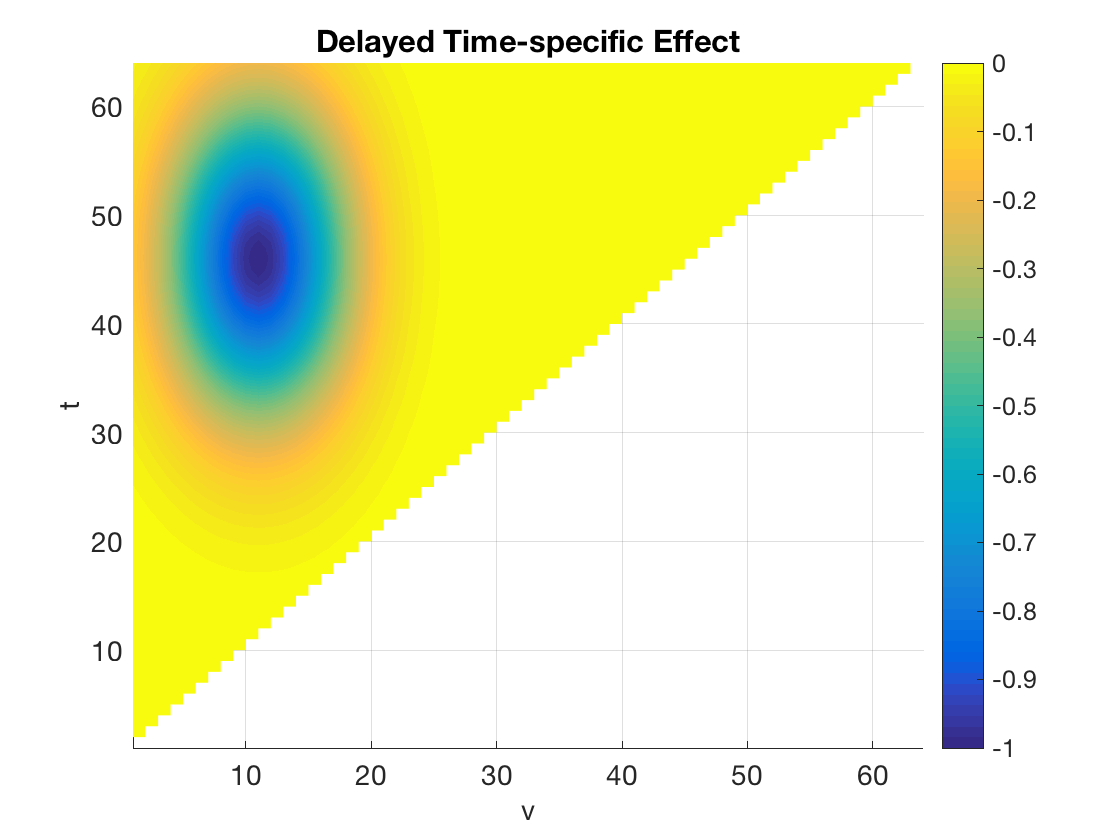}
		\caption{Historical surface scenarios for $T = 64$ measurements. This figure appears in color in the electronic version of this article. \label{f:true}}
	\end{figure}

\begin{table}
\caption{RMISEs, average percent of energy preserved in $\bX^{W_P}$ (PE), and computation time for the simulation settings lagged ($L$), cumulative ($C$), time-specific ($T$), and delayed time-specific ($D$). Table values represent averages taken over 200 simulated datasets. RC denotes retained coefficients. Time is averaged over all simulated datasets, over all settings.\label{t:rmise}}
\begin{center}
\begin{tabular}{llllcccclclc}
\hline
\hline
 \multirow{2}{*}{N} & \multirow{2}{*}{T} & \multirow{2}{*}{RC} &  & \multicolumn{4}{c}{RMISE} &  & \multirow{2}{*}{PE} & & \multirow{2}{*}{Time (s)} \\
  \cline{5-8} 
& & & & $\bbeta_L$ & $\bbeta_C$ & $\bbeta_T$ & $\bbeta_D$ & &  & & \\
\hline
50 & 64 & 25\% &  & 0.015 & 0.012 & 0.095 & 0.045 &  & 73.67\% &  & 516.2 \\
 &  & 50\% &  & 0.018 & 0.017 & 0.056 &  0.017 &  & 88.42\% &  & 609.6 \\
 \cline{3-12}
 & 128 & 25\% &  & 0.010 & 0.008 & 0.040 &  0.008 &  & 74.91\% &  & 1035.7 \\
\hline
200 & 64 & 25\% &  & 0.009 & 0.007 & 0.088 & 0.009 &  & 73.55\% &  & 515.6 \\
 & & 50\% &  & 0.010 & 0.008 & 0.045 & 0.008 &  & 88.34\% &  & 615.5 \\
 \cline{3-12}
 & 128 & 25\% &  & 0.004 & 0.004 & 0.036 & 0.003 &  & 75.63\% &  & 1032.1 \\
\hline
\hline
\end{tabular}
\end{center}
\end{table}




To evaluate each setting, we calculate the root mean integrated squared error (RMISE) and determine coverage probabilities for both point-wise and joint credible intervals. We also graphically explore the properties of the BFDR procedure. For each ``true'' historical surface $\bbeta$, we generate $N$ $\bx_i$ curves from a mean zero Gaussian Process with a first-order auto-regressive (AR1) covariance structure. We base the variance and correlation parameters of the AR1 covariance off of the PM$_2.5$ data from the Journeyman data, setting $\sigma^2_{AR, X} = 3.5$ and $\rho_X = 0.75$. Next we generate within-subject error functions, $\be_i$, from a separate mean zero Gaussian Process with an AR1 covariance structure. Once again we base the parameters of the covariance matrix off of the Journeyman data, $\sigma^2_{AR,E} = 0.1$ and $\rho_E = 0.5$. We then simulate the outcome functions using $\by_i = \bx_i\bbeta + \be_i$. For each setting, we repeat this data generation process 200 times and obtain 2000 posterior samples, discarding the first 1000, for each simulated dataset. For each simulated dataset we perform a $J = 3$ level DWPT on both the outcome and predictor using Daubechies wavelets with 3 vanishing moments. All computation is done using MATLAB version R2017a on a desktop with a 3.2 GHz Intel Core i5 processor and 16 GB of memory.

\begin{figure}
		\centering
		\includegraphics[width = 3in]{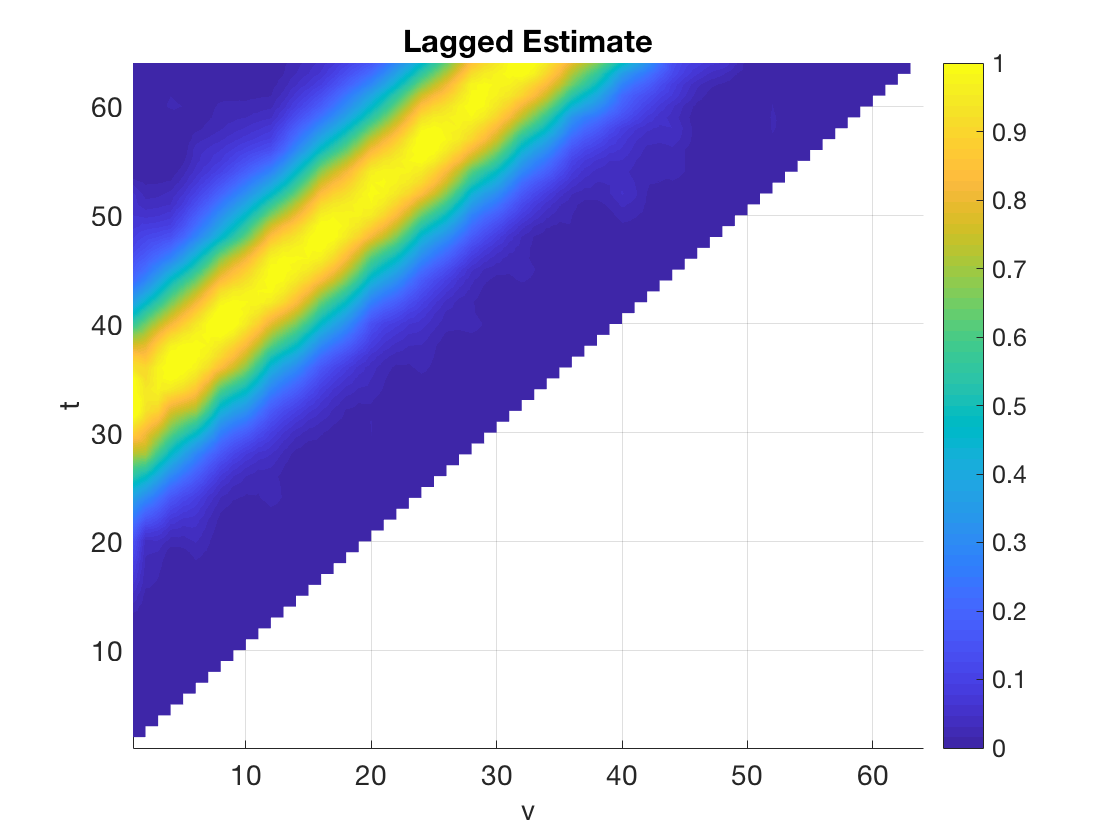}
		\includegraphics[width = 3in]{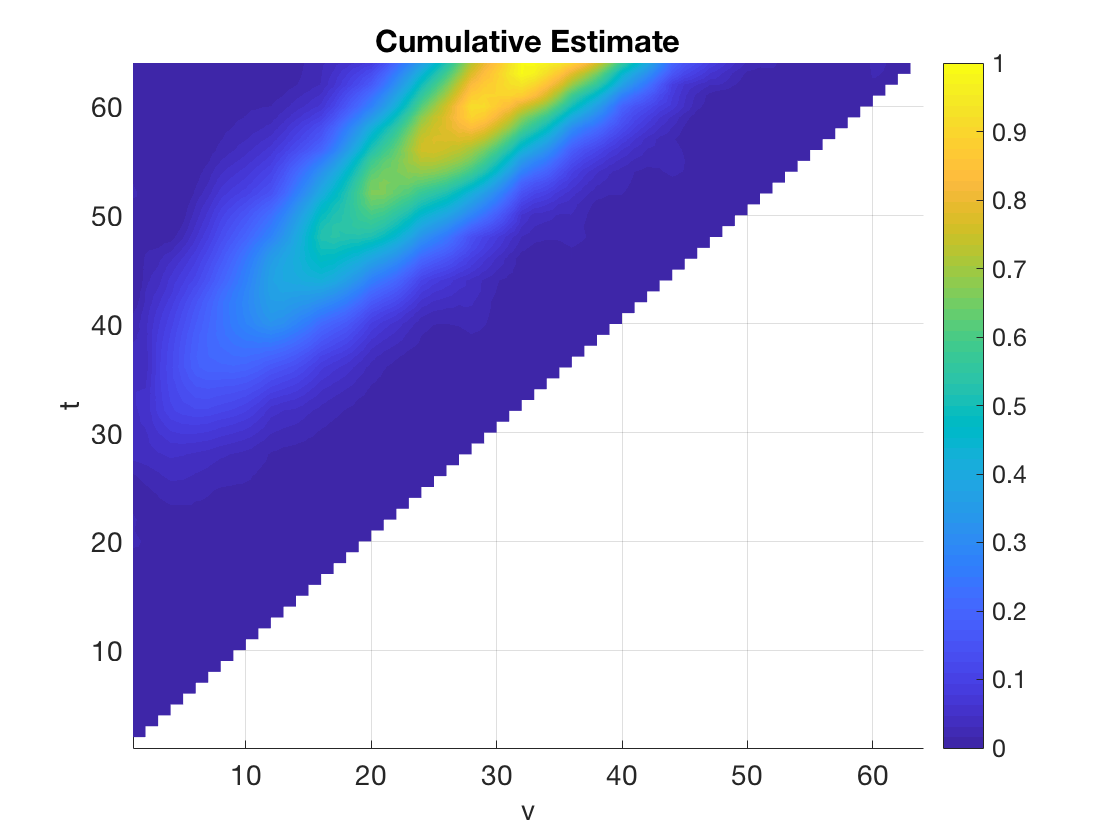}
		\includegraphics[width = 3in]{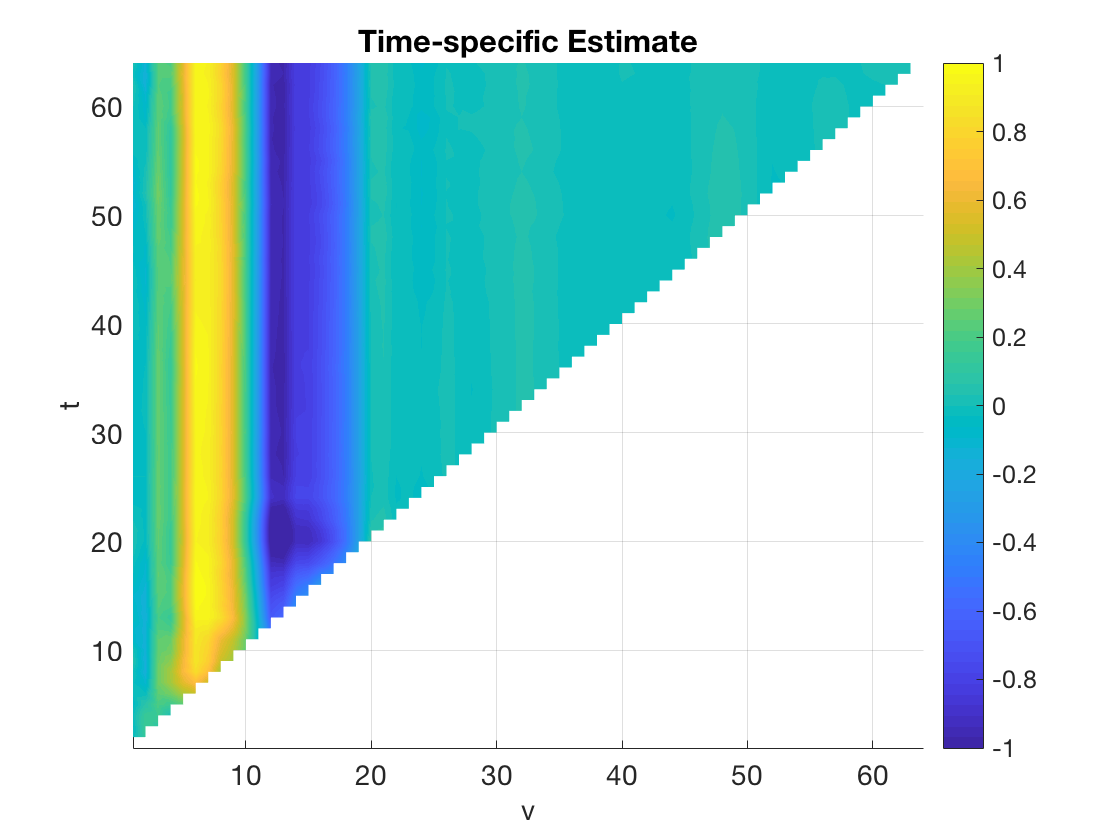}
		\includegraphics[width = 3in]{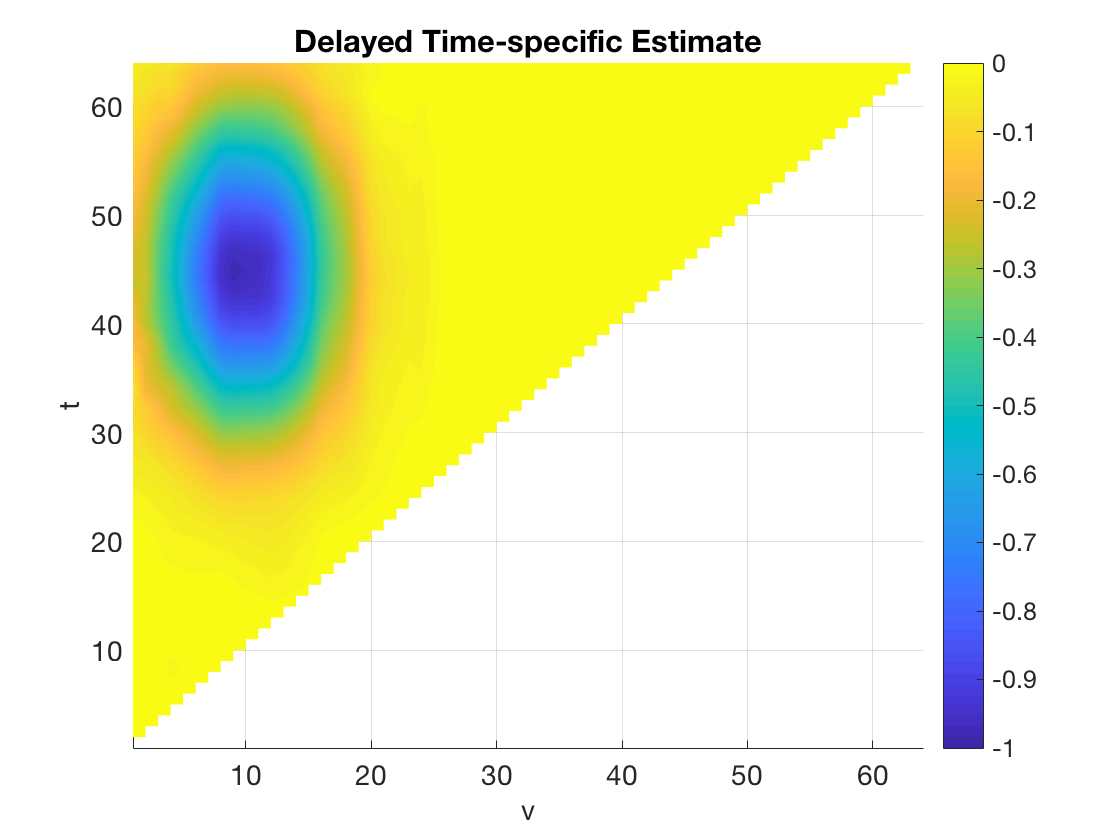}
		\caption{Estimated surfaces from one simulation setting with $N = 50$ subjects, $T = 64$ measurements, 25\% retained columns of $\bX^{W_P}$, and near average RMISE. This figure appears in color in the electronic version of this article. \label{f:est}}
	\end{figure}

Table~\ref{t:rmise} presents RMISE averaged across all 200 simulated datasets for each setting under consideration while Figure~\ref{f:est} displays a single estimated surface with near average RMISE for the case when $N = 50$, $T = 64$ and only 25\% of the $\bX^{W_P}$ coefficients are retained. From Table~\ref{t:rmise}, we see that increasing either the sample size or the percent of retained coefficients tends to decrease RMISE, although all settings produce similar results.
Denser sampling also decreases the RMISE. Regardless of sample size or sampling density, retaining 25\% of the $X$-space wavelet-packet coefficients preserves roughly 75\% of the energy while retaining 50\% of coefficients preserves 88\% of the energy in $X(v)$. To calculate average preserved energy, we find it first for each simulated subject and then average across all subjects. From Figure~\ref{f:est}, we see that a single simulated dataset with near average RMISE accurately estimates the true relationship, even when only retaining an average of 75\% of the energy in the predictor function.

\begin{figure}
		\centering
		\includegraphics[width = 3in]{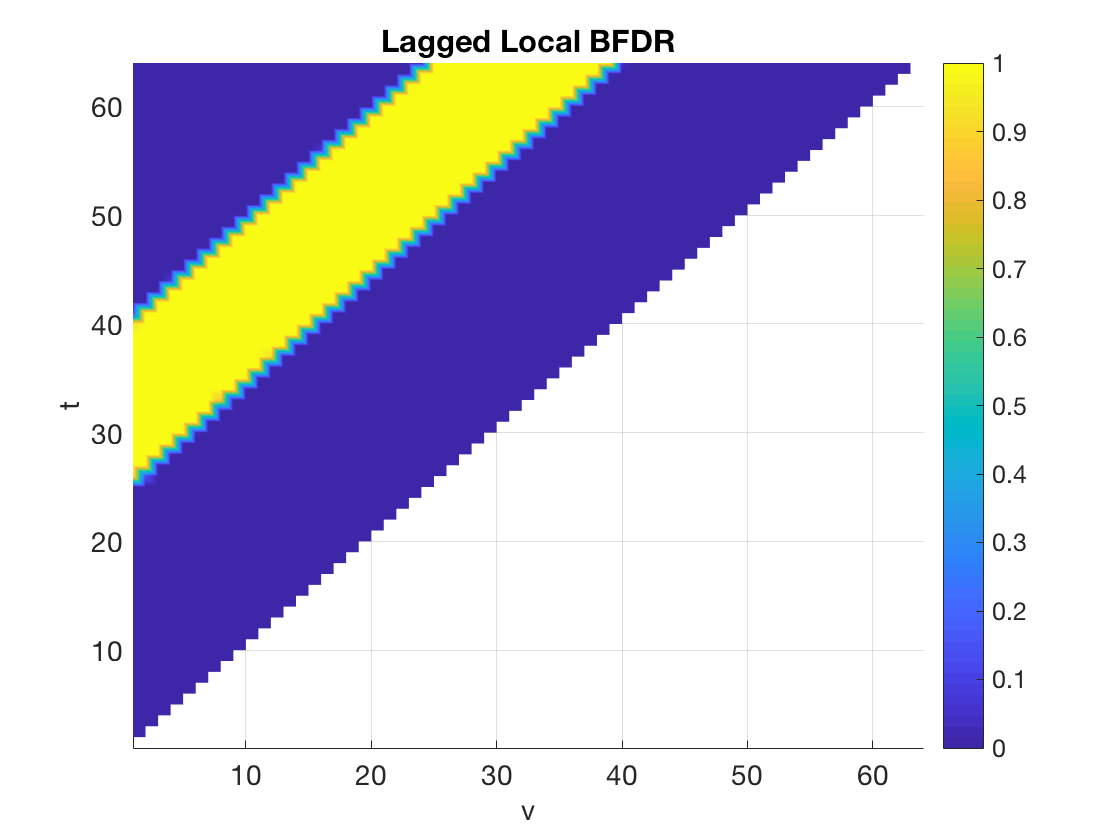}
		\includegraphics[width = 3in]{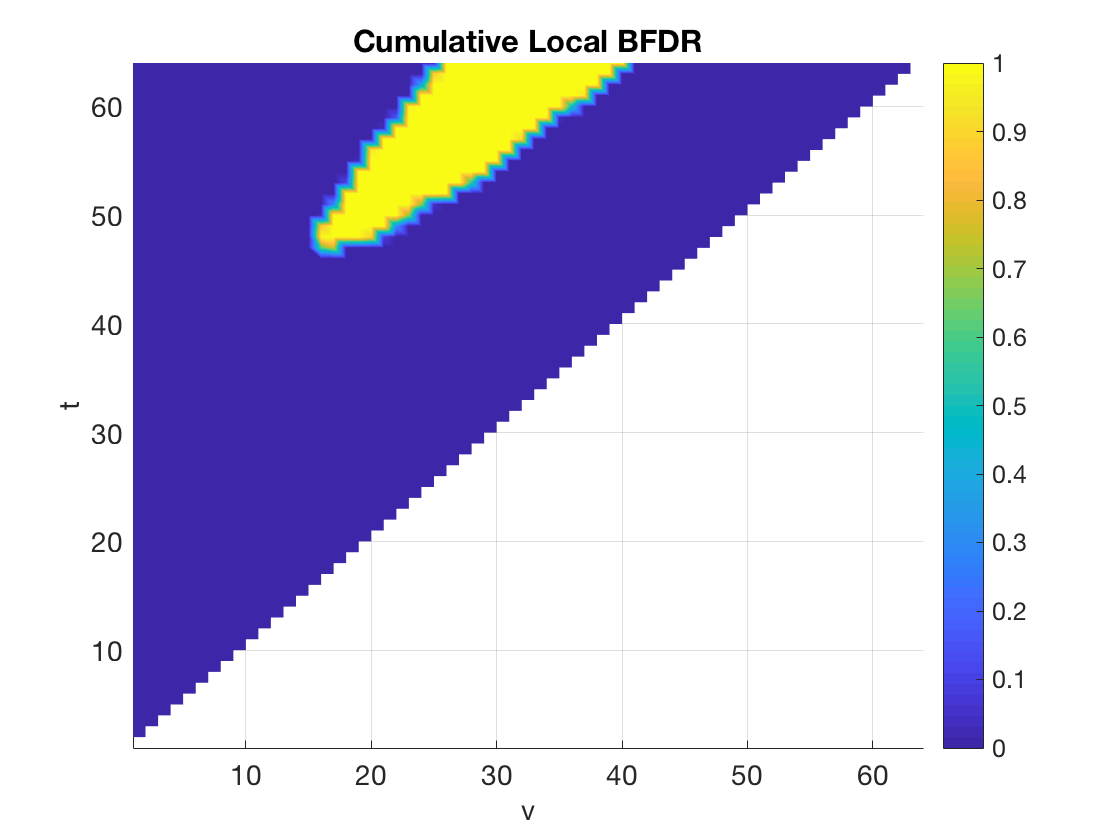}
		\includegraphics[width = 3in]{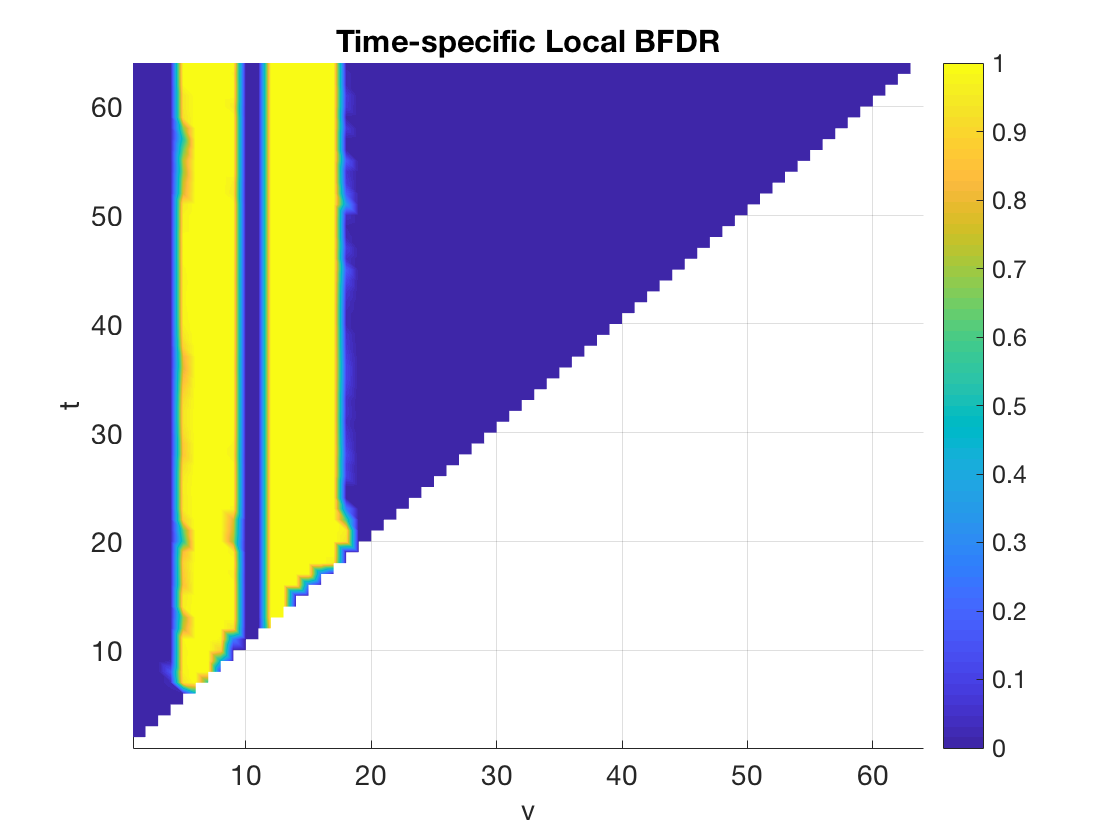}
		\includegraphics[width = 3in]{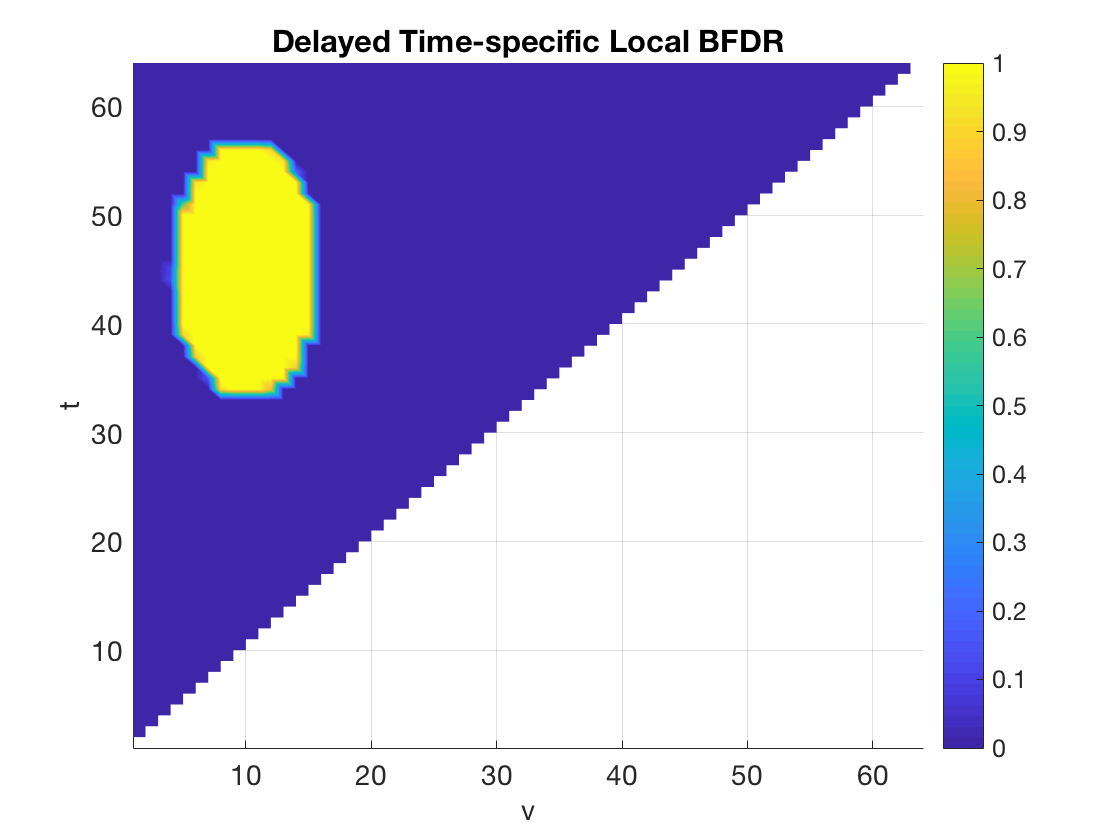}
		\caption{Local BFDR, $P_B(v,t)$, values using $\delta = 0.50$ for the estimated surfaces from one simulated data set with $N = 50$ subjects, $T = 64$ measurements, and near average RMISE. This figure appears in color in the electronic version of this article. \label{f:pst50}}
	\end{figure}

	We assess inference using Table~\ref{t:cover} and Figure~\ref{f:pst50}. Table~\ref{t:cover} contains mean point-wise coverage probabilities, calculated by first averaging over the historical surface and then over all 200 simulated datasets. We calculate coverage for both the point-wise credible interval and the joint credible interval. From Table~\ref{t:cover}, we see that in general the joint credible intervals provide higher coverage for each setting. Coverage also tends to increase, regardless of interval type, as the percent of retained coefficients increase, though in some settings the differences are negligible. For some settings, coverage of the joint intervals is affected by sampling density and sample size. However, for the joint intervals, coverage is above the nominal level for most settings. This is not the case for the point-wise intervals where coverage sits below nominal level, in some bases considerably below. All intervals are at the 95\% level.
	
\begin{table}
\caption{Point-wise and joint credible interval coverage probabilities for the simulation settings lagged ($L$), cumulative ($C$), time-specific ($T$), and delayed time-specific ($D$). All intervals are at the 95\% level. Table values represent averages taken over 200 simulated datasets. RC denotes retained coefficients. \label{t:cover}}
\begin{center}
\begin{tabular}{llllcccclcccc}
\hline
\hline
  \multirow{2}{*}{N}  & \multirow{2}{*}{T} & \multirow{2}{*}{RC} & & \multicolumn{4}{c}{Point-wise Interval} &  & \multicolumn{4}{c}{Joint Interval} \\
  \cline{5-8} \cline{10-13}
  &  &  &   & $\bbeta_L$ & $\bbeta_C$ & $\bbeta_T$ & $\bbeta_D$ & & $\bbeta_L$ & $\bbeta_C$ & $\bbeta_T$ & $\bbeta_D$  \\
\hline
50 & 64 & 25\% &  & 0.811 & 0.870 & 0.524 & 0.565 &  & 0.993 & 0.997 & 0.770 & 0.778 \\
 &  & 50\% &  & 0.923 & 0.942 & 0.826 & 0.934 &  & 0.999 & 0.999 & 0.984 & 0.999  \\
\cline{3-13}
 & 128 & 25\% &  & 0.855 & 0.885 & 0.767 &  0.940 &  & 1.000 & 1.000 & 0.978 & 1.000  \\
\hline
200 & 64 & 25\% &  & 0.724 & 0.784 & 0.452 & 0.735 &  & 0.957 & 0.986 & 0.673 & 0.957 \\
 &  & 50\% &  & 0.865 & 0.890 & 0.718 & 0.930 &  & 0.999 & 0.999 & 0.933 & 1.000 \\
\cline{3-13}
 & 128 & 25\% &  & 0.884 & 0.881 & 0.590 & 0.897 &  & 0.999 & 0.999 & 0.906 & 1.000 \\
\hline
\hline
\end{tabular}
\end{center}
\end{table}
	
	Figure~\ref{f:pst50} shows the local BFDR values using a cut-off of $\delta = 0.5$ which is equal to half the largest signal, in absolute value, from each surface. Each plotted value corresponds to the posterior probability that $|\beta(v,t)| > 0.5$ for each surface from Figure~\ref{f:est}, that is for one simulated dataset with near average RMISE. We see that using this $\delta$-intensity change, the BFDR highlights the features of each setting with elevated (or depressed) association. In the Supplementary Material, we consider one other value of $\delta$, which, while smaller, produces similar results. Choice of a specific $\delta$ is data-dependent, but several different values should be considered. 

	Figures~\ref{f:est} and~\ref{f:pst50} present the results from the simulation for $N = 50$, $T = 64$, and retaining 25\% of the columns of $\bX^{W_P}$. Graphical results from simulated datasets with near average RMISE for the other settings produce similar results. Consequently, we present limited additional BFDR results in the Supplementary Material. Specifically, we only show BFDR graphs for all remaining 25\% retained coefficients and $T = 64$ settings for $\delta = 0.25$ and 0.50.

\section{Analysis of Journeyman Boilermaker Data}
\label{s:app}

\cite{Harezlak2007} and \cite{Cavallari2008} analyze data from 14 Journeyman boilermakers relating their five-minute SDNN to five-minute PM$_{2.5}$ levels over the course of the workday. Exposure to PM$_{2.5}$ came from two sources: residual oil fly-ash, a by-product of the manufacturing process, and cigarette smoke during mandatory breaks. \cite{Harezlak2007} found elevated levels association during the morning hours that corresponded to mandatory breaks which were then followed by depressions in the surface estimated over the trapezoidal region. To try to better estimate these spikes in association, we focus our analysis on the first three hours of the workday beginning at 8:30am and going until 11:30am. Further, we estimate joint intervals and calculate the BFDR to determine the significance of these peaks and troughs.

	\begin{figure}
		\centering
		\includegraphics[width = 3in]{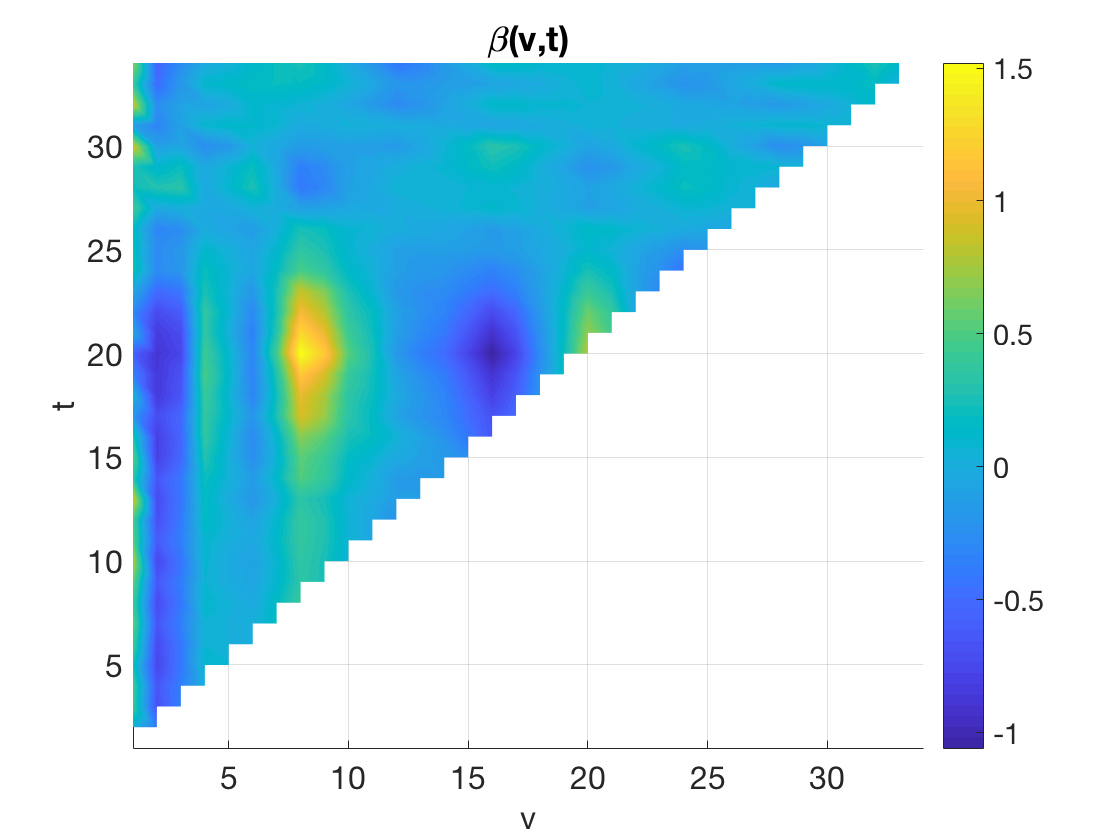}
		\caption{Estimated surface of association between $\log$(SDNN) and $\log$(PM$_{2.5}$) in Journeyman boilermakers during the morning, 8:30am to 11:30am. This figure appears in color in the electronic version of this article. \label{f:jourest}}
	\end{figure}
	
	Prior to analysis, we log transform and center and scale both SDNN and PM$_{2.5}$. Thus changes in the estimated surface correspond to one unit changes in $\log($PM$_{2.5})$ and result in changes in $\log$(SDNN). To investigate the morning hours only, we take $T = V = 34$ measurements. As in the simulation, we take $J = 3$ levels of decomposition for both wavelet-packet transformations and use Daubechies wavelets with 3 vanishing moments. Given the results of the simulation, we retain only the first two levels of wavelet-packet coefficients, resulting in only 25\% of the columns of $\bX^{W_P}$ being kept. We let this model run longer than the simulated settings, taking 2000 posterior samples after a burnin of 2000. We monitor convergence with running mean and trace plots to ensure all chains converged (see Supplementary Material).
	
\begin{figure}
		\centering
		\includegraphics[width = 3in]{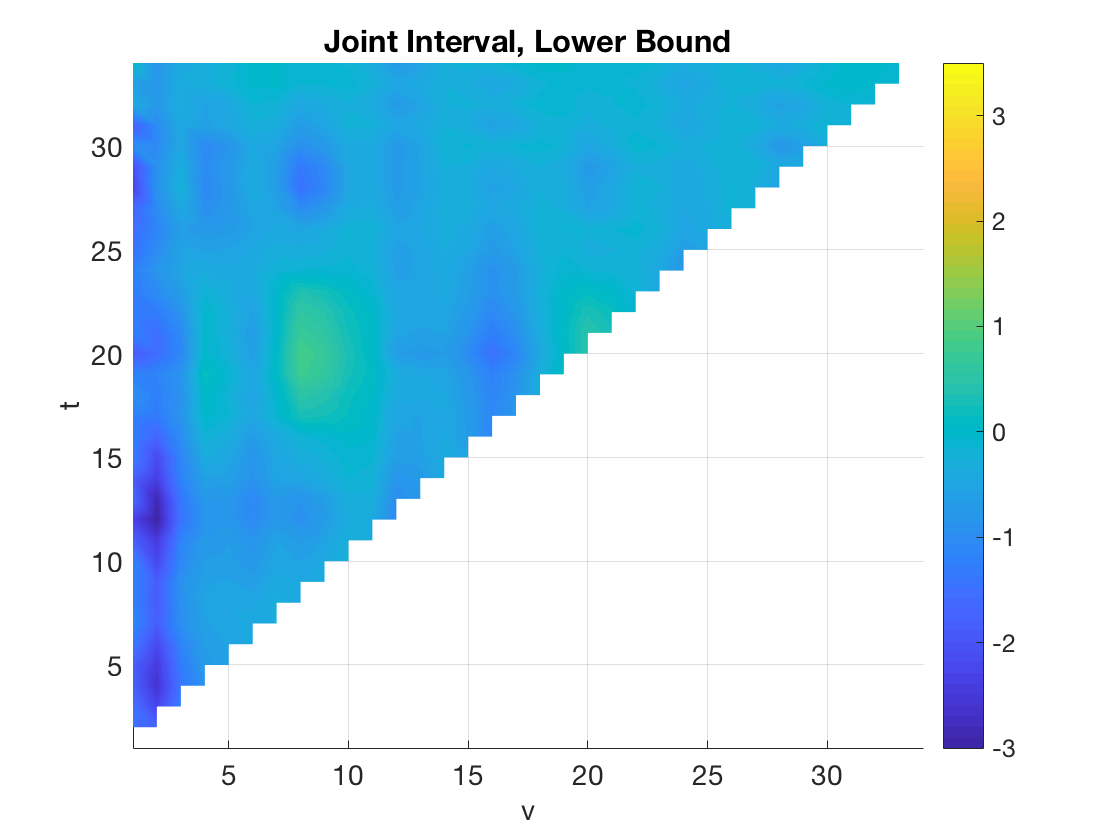}
		\includegraphics[width = 3in]{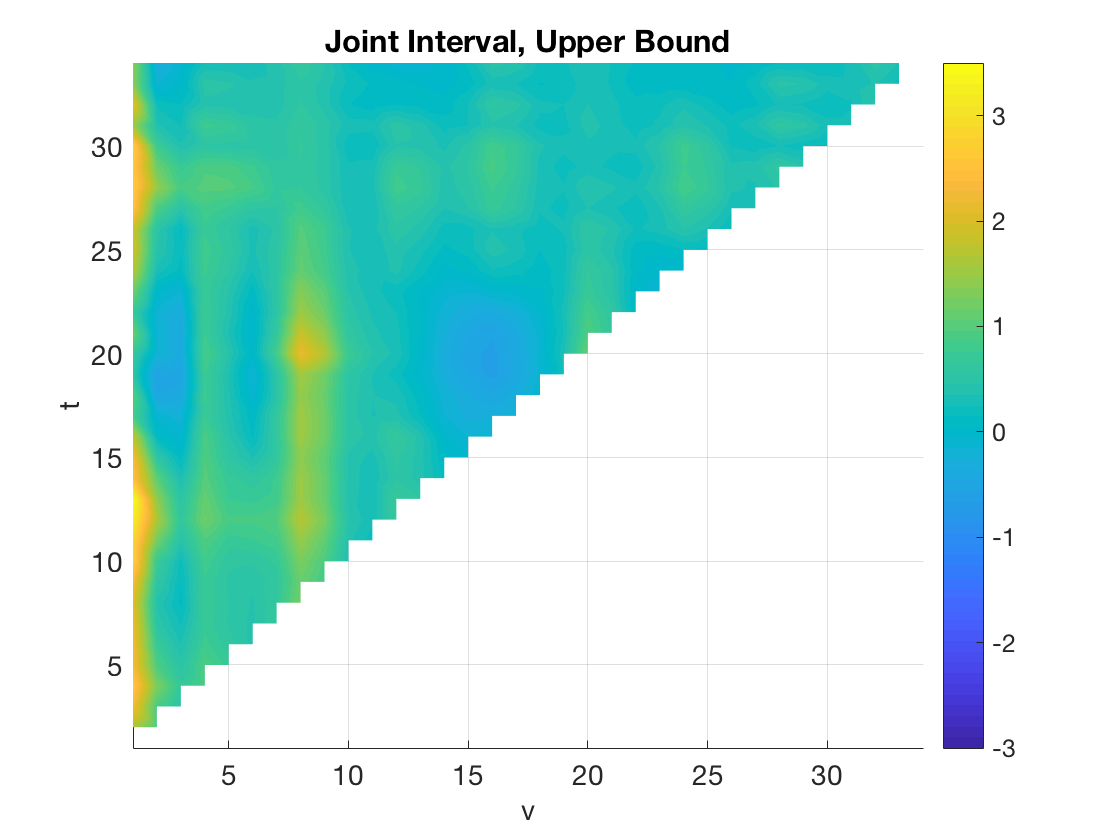}
		\caption{Lower (left) and upper (right) bounds of the joint interval for the estimated surface of association between $\log$(SDNN) and $\log$(PM$_{2.5}$) in Journeyman boilermakers. This figure appears in color in the electronic version of this article. \label{f:jourint}}
	\end{figure}

	Figure~\ref{f:jourest} contains the posterior estimated historical surface for the Journeyman data whereas Figure~\ref{f:jourint} presents the lower and upper bounds of the joint credible intervals. From Figure~\ref{f:jourest}, moving across the $v$-axis from left to right, we see a pattern of time-specific depressions followed by elevations, which culminate in a large peak around $v = 8$ and $t = 20$. This suggests that an exposure 40 minutes into the day resulted in elevated SDNN that was sustained until, and peaked  around, 100 minutes. After this peak, there is another depression that occurs around $v = 16$ and $t = 20$. These patterns are consistent with the analysis in \cite{Harezlak2007}.

	The joint bands experience edge effects along the $t$-axis, which somewhat distort the scale of the intervals in Figure~\ref{f:jourint}. However, from this figure, we are able to see the elevation of the lower bound around the major peak around $v = 8$ and $t = 20$ as well as the depression around $v = 16$ and $t = 20$. The ridge that runs along $v = 3$ in Figure~\ref{f:jourest} also appears to have an elevated lower bound. In the Supplementary Material, we present the point-wise intervals which are smaller in width, but exhibit similar features to the joint intervals. The intervals suggest regions of both increased and decreased association between SDNN and PM$_{2.5}$ throughout the morning. To more accurately pinpoint these regions, we turn to the BFDR procedure.

	\begin{figure}
		\centering
		\includegraphics[width = 3in]{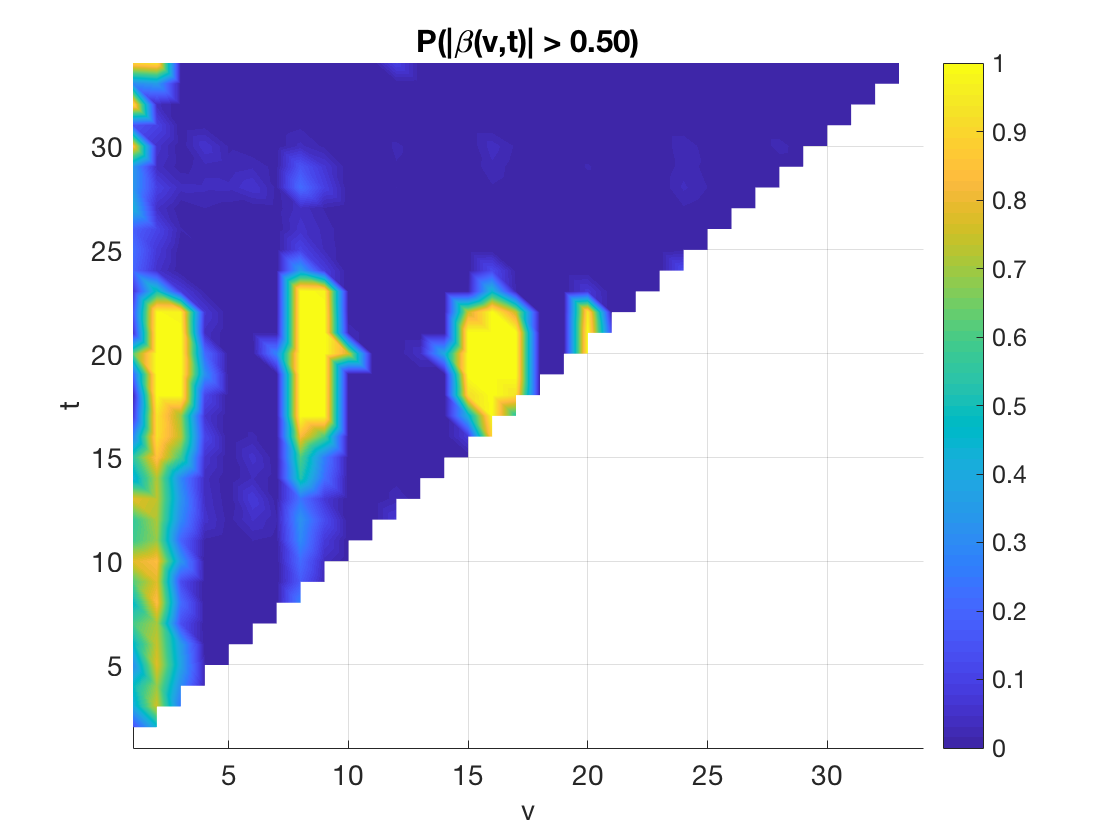}
		\includegraphics[width = 3in]{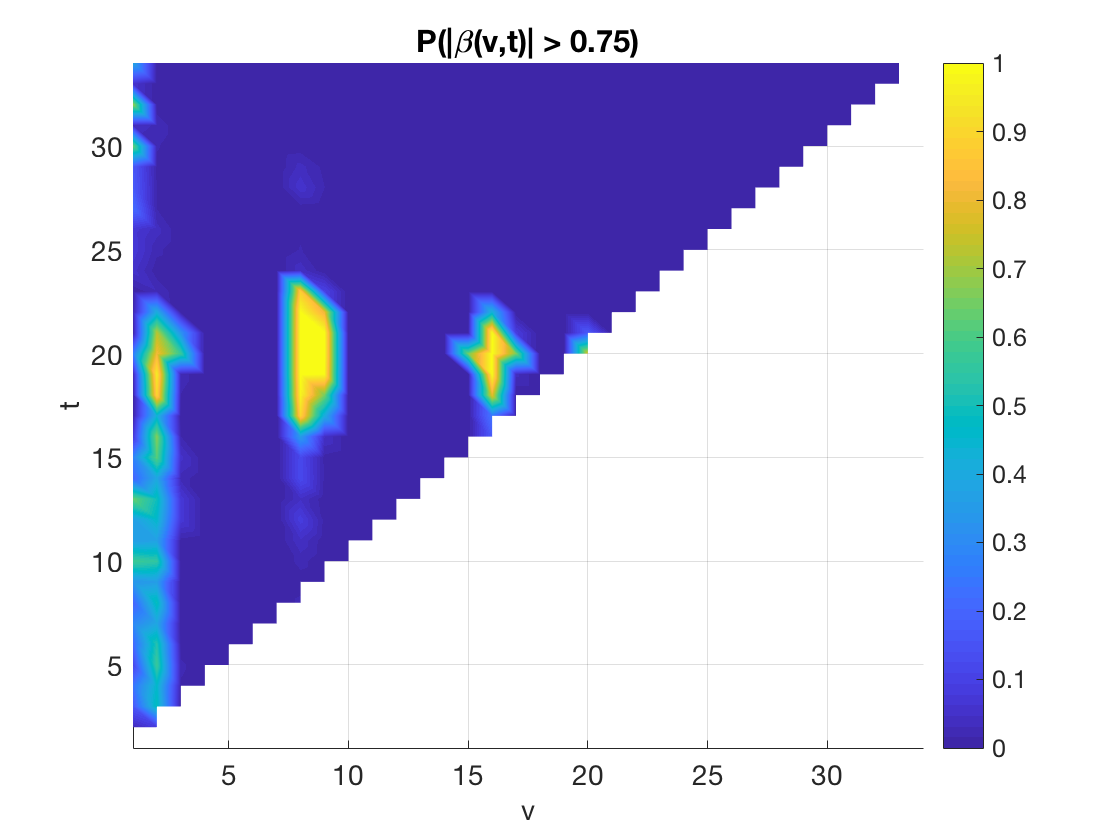}
		\includegraphics[width = 3in]{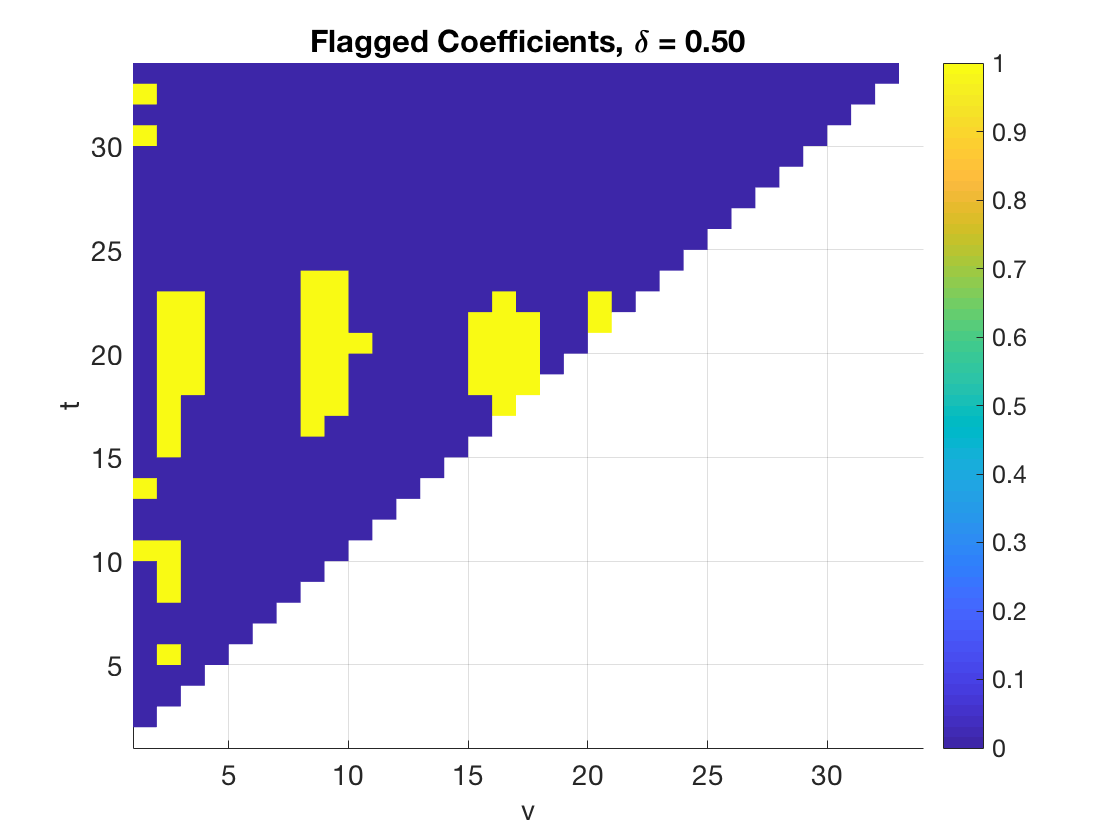}
		\includegraphics[width = 3in]{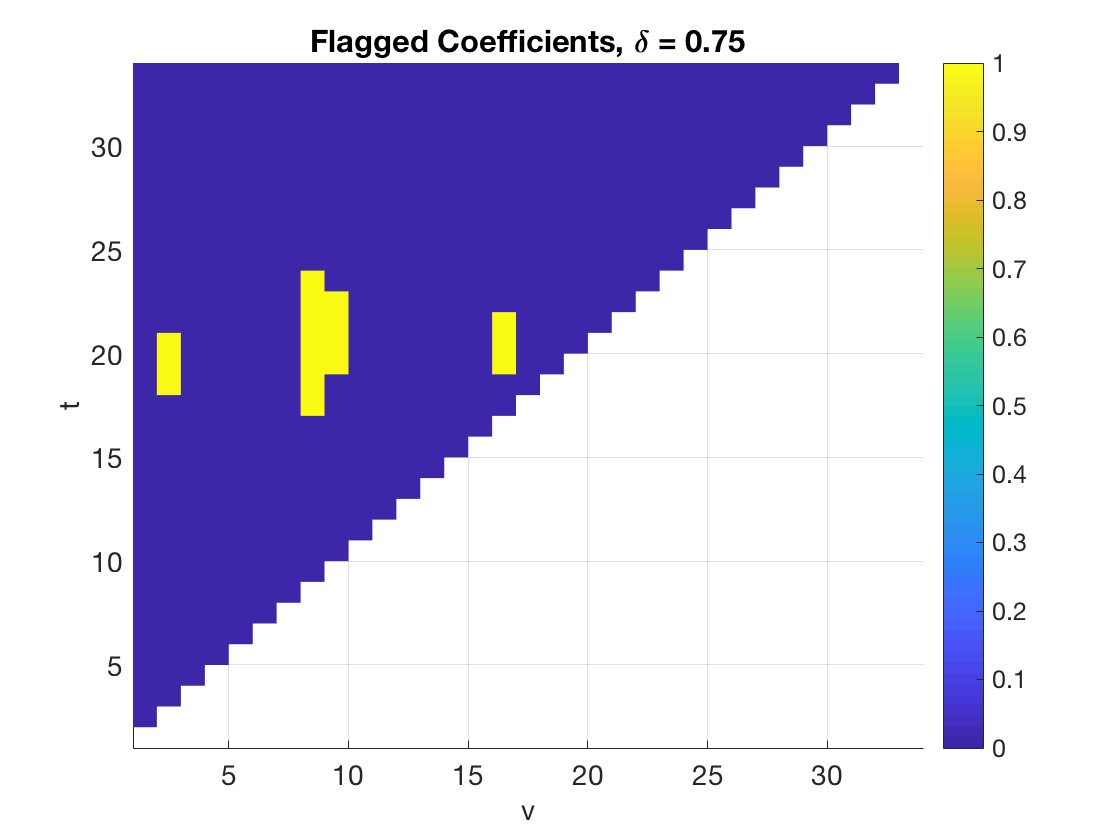}
		\caption{Local BFDR (top row) and flagged significant coefficients (bottom row) for the estimated surface of association between $\log$(HRV) and $\log$(PM$_{2.5}$) in Journeyman boilermakers. This figure appears in color in the electronic version of this article. \label{f:jourfdr}}
	\end{figure}

Figure~\ref{f:jourfdr} presents the local BFDR values, $P_B(v,t)$, for each coefficient (top row) as well as the flagged coefficients from the set $\psi_{\alpha}$ (bottom row) using $\alpha = 0.05$. The columns correspond to two different $\delta$ intensity cut-offs representing corresponding changes in $\log$(SDNN) of 0.5 and 0.75. The lower cut-off value highlights significant coefficients in the regions where we see the strongest effect sizes from Figure~\ref{f:jourest}. These regions do not extend from the diagonal nor to the edge of the surface, suggesting a series of delayed time-specific effects that result from exposures within the first 100 minutes of the workday. Some of these effects are strong, as evidenced by the sustained significance at the higher $\delta$ level. In the Supplementary Material, we present additional results which shows the main peak around $v = 8$ and $t = 20$ is significant at an intensity change of $\delta = 1.00$.

\section{Discussion}
\label{s:disc}

One of the difficulties of estimating a historical functional effect is in maintaining the historical constraint. Previous authors have used tent shaped basis functions to achieve the constraint. When using wavelets coefficients from a DWT as a basis function, constraining the surface is difficult due to the lack of a clear relationship between the wavelet-space coefficients and the time domain. However, wavelet-packet coefficients from a DWPT have an exploitable relationship between their location indices and the original time scale that allows us to constrain the surface in the wavelet-packet space and maintain that constraint when projecting back to the data-space. In this work, we show that when performing the DWPT on both $y(t)$ and $x(v)$, we can use wavelet-packet basis functions, for a given choice of mother wavelet, to model historical effects.

The current literature on HFLMs is limited, largely focusing on the estimation of surface effects and the determination of model fit criterion. Most methods implement spline-based basis functions and, while one method does allow for different basis expansions, the authors only present results for spline-based models. Further, inference procedures are not discussed in the existing literature. To the best of our knowledge, the model we present here represents the first work in wavelet-based modeling of historical effects as well as the first Bayesian HFLM. Additionally, our method employs a novel use of wavelet-packets, which have not previously been used in functional regression models, to constrain the surface of estimation. Finally, we perform inference using two established multiplicity adjusted Bayesian inferential procedures for describing the uncertainty in the surface estimate through the use of joint credible intervals and identifying significant regions of coefficients using the BFDR.

We demonstrate in simulation that the wavelet-packet HFLM can accurately estimate several realistic historical surface settings. In particular, we show that regardless of the percent of $\bX^{W_P}$ coefficients we retain, the wavelet-packet HLFM has good RMISE levels that are similar across sample size and sampling density. Further, the joint credible intervals used in \cite{Meyer2015} for the wavelet-based FFR outperform point-wise credible intervals in terms of average coverage. And while their coverage is larger than the nominal, we prefer this to the alternative. Thus we suggest the use of the joint intervals for quantifying uncertainty in the wavelet-packet HFLM. To identify regions of significant coefficients, we implement the BFDR to show that for two reasonable choices of $\delta$ relative to the data and max signal, our model can identify regions of association in the surface. Finally, we apply the proposed model to analyze data on the association between HRV and PM$_{2.5}$ exposure in a panel of journeymen, focusing on the first three hours of exposure during the workday. Using the wavelet-packet HFLM, we are able to not only estimate regions of association but clearly identify them as representing significant changes in SDNN using the BFDR.

\bigskip
\begin{center}
{\large\bf SUPPLEMENTARY MATERIAL}
\end{center}

\begin{description}

\item[Supplementary Material to Bayesian HFLM with Wavelet-packets:] Contains additional results from the simulation and application. (pdf)

\item[MATLAB Code:] Available at \url{https://github.com/markjmeyer/WPHFLM}.

\end{description}


\begin{thebibliography}{}

\bibitem[\protect\citeauthoryear{Cavallari et al.}{2008}]{Cavallari2008} Cavallari, J. M., Fang, S. C., Eisen, E. A., Schwartz, J., Hauser, R., and Herrick, R. F. (2008). Time Course of Heart Rate Variability Decline Following Particulate Matter Exposures in an Occupational Cohort. \textit{Inhalation Toxicology} \textbf{20,} 415--422.

\bibitem[\protect\citeauthoryear{Harezlak et al.}{2007}]{Harezlak2007} Harezlak, J., Coull, B. A., Laird, N. M., Magari, S. R., and Christiani, D. C. (2007). Penalized solutions to functional regression problems. \textit{Computational Statistics \& Data Analysis} \textbf{51,} 4911--4925.

\bibitem[\protect\citeauthoryear{Ivanescu et al.}{2015}]{Ivanescu2015} Ivanescu, A.E., Staicu, A.-M., Scheipl, F., and Greven, S. (2015). Penalized function-on-function regression. \textit{Computational Statistics} \textbf{30,} 539--568.

\bibitem[\protect\citeauthoryear{Kim et al.}{2018}]{Kim2018} Kim, J. S., Staicu, A.-M., Maity, A., Carroll, R. J., and Ruppert, D. (2018). Additive Function-on-Function Regression. \textit{Journal of Computational and Graphical Statistics} \textbf{27,} 234--244.

\bibitem[\protect\citeauthoryear{Kim, \c{S}ent\"{u}rk, and Li}{2011}]{Kim2011} Kim, K., \c{S}ent\"{u}rk, D., and Li, R. (2011). Recent history functional linear models for sparse longitudinal data. \textit{Journal of Statistical Planning and Inference} \textbf{141,} 1554--1566.

\bibitem[\protect\citeauthoryear{Magari et al.}{2001}]{Magari2001} Magari, S. R., Hauser, R., Schwartz, J., Williams, P. L., Smith, T. J., and Christiani, D. C. (2001). Association of Heart Rate Variability With Occupational and Environmental Exposure to Particulate Air Pollution. \textit{Circulation} \textbf{104,} 986--991.

\bibitem[\protect\citeauthoryear{Malfait and Ramsay}{2003}]{MalfaitRamsay2003} Malfait, N. and Ramsay, J. O. (2003). The historical functional linear model. \textit{The Canadian Journal of Statistics} \textbf{31,} 115--128.

\bibitem[\protect\citeauthoryear{Malloy et al.}{2010}]{Malloy2010} Malloy, E. J., Morris, J. S., Adar, S. D., Suh, H., Gold, D. R., and Coull, B. A. (2010). Wavelet-based functional linear mixed models: an application to measurement error-corrected distributed lag models. \textit{Biostatistics} \textbf{11,} 432--452.

\bibitem[\protect\citeauthoryear{Meyer et al.}{2015}]{Meyer2015} Meyer, M. J., Coull, B. A., Versace, F., Cinciripini, P., and Morris, J. S. (2015). Bayesian Function-on-Function Regression for Multi-Level Functional Data. \textit{Biometrics} \textbf{71,} 563--574.

\bibitem[\protect\citeauthoryear{Misiti et al.}{2007}]{Misiti2007} Misiti, M., Misiti, Y., Oppenheim, G., and Poggi, J.-M. (2007).\textit{Wavelets and their Applications}. ISTE Ltd.

\bibitem[\protect\citeauthoryear{Morris}{2015}]{Morris2015} Morris, J. S. (2015). Functional regression. \textit{Annual Reviews of Statistics and Its Applications} \textbf{2,} 321--359.

\bibitem[\protect\citeauthoryear{Morris et al.}{2008}]{MorrisEtAl2008} Morris, J. S., Brown, P. J., Herrick, R. C., Baggerly, K. A., and Coombes, K. R. (2008). Bayesian Analysis of Mass Spectrometry Proteomic Data Using Wavelet-Based Functional Mixed Models.  \textit{Biometrics} \textbf{64,} 479--489.

\bibitem[\protect\citeauthoryear{Morris and Carroll}{2006}]{MorrisCarroll2006} Morris, J. S. and Carroll, R. J. (2006). Wavelet-based functional mixed models.  \textit{Journal of the Royal Statistical Society, Series B} \textbf{68,} 179--199.

\bibitem[\protect\citeauthoryear{Nason}{2008}]{Nason2008} Nason, G. P. (2008). \textit{Wavelet Methods in Statistics with R}. Springer.

\bibitem[\protect\citeauthoryear{Percival and Walden}{2000}]{PercivalWalden2000} Percival, D. B. and Walden, A. T. (2000). \textit{Wavelet Methods for Time Series Analysis}. Cambridge University Press.

\bibitem[\protect\citeauthoryear{Pomann et al.}{2016}]{Pomann2016} Pomann, G.-M., Staicu, A.-M., Lobaton, E. J., Mejia, A. F., Dewey, B. E., Reich, D. S., Sweeney, E. M., and Shinohara,  R. T. (2016). A lag functional linear model for prediction of magnetization transfer ratio in multiple sclerosis lesions. \textit{Annals of Applied Statistics} \textbf{10,} 2325--2348.

\bibitem[\protect\citeauthoryear{Ruppert, Wand, and Carroll}{2003}]{Ruppert2003} Ruppert, D., Wand, M. P., and Carroll, R. J. (2003). \textit{Semiparametric Regression}. Cambridge University Press.

\bibitem[\protect\citeauthoryear{Scheipl and Greven}{2016}]{Scheipl2016} Scheipl, F. and Greven, S. (2016). Identifiability in penalized function-on-function regression models. \textit{Electronic Journal of Statistics} \textbf{10,} 495--526.

\bibitem[\protect\citeauthoryear{Scheipl, Staicu, and Greven}{2015}]{Scheipl2015} Scheipl, F., Staicu, A.-M., and Greven, S. (2015). Functional Additive Mixed Models. \textit{Journal of Computational and Graphical Statistics} \textbf{24,} 477--501.

\end{thebibliography}
\end{document}